\pdfoutput=1 
\documentclass[10pt,conference,letterpaper]{IEEEtran}
\ifCLASSINFOpdf
\else
\fi
\usepackage{settings}

\begin{document}

\title{
{\Large \bf Cyber-Physical Interference Modeling for\\
\vspace*{-0.15in}  Predictable Reliability of Inter-Vehicle Communications 
\vspace*{-0.2in}
} 
\ifthenelse{\boolean{short} \OR \boolean{anony}}
{}
{ \\ {\normalsize (Technical Report: WSU-CS-DNC-TR-17-01) } }
}
%
%
%
%

\author{
\ifthenelse{\boolean{anony}}
{
\ifthenelse{\boolean{short}}
{\vspace*{-0.3in} Paper \#157 \vspace*{-0.4in} }
{Full-length version of the ICNP'17 submission \#157 }
}
{ 
\IEEEauthorblockN{Chuan Li$^\star$, Hongwei Zhang$^\star$, Jayanthi Rao$^\dagger$, 
Le Yi Wang$^\star$, George Yin$^\star$ }
\IEEEauthorblockA{$^\star$Wayne State University, \{chuan,hongwei,lywang,gyin\}@wayne.edu}
\IEEEauthorblockA{$^\dagger$Ford Research, jrao1@ford.com 
\vspace*{-0.3in}
}

} 
} 

\maketitle



\begin{abstract}
Predictable inter-vehicle communication reliability is a basis for the paradigm shift from the traditional single-vehicle-oriented safety and efficiency control to networked vehicle control. The lack of predictable interference control in existing mechanisms of inter-vehicle communications, however, makes them incapable of ensuring predictable communication reliability. For predictable interference control, we propose the Cyber-Physical Scheduling (CPS) framework that leverages the PRK interference model and addresses the challenges of vehicle mobility
\ifthenelse{\boolean{ieee10pager}}
{to PRK-based scheduling. In particular, }
{and broadcast to PRK-based scheduling. To address vehicle mobility, }
    CPS leverage physical locations of vehicles to define the gPRK interference model, a geometric approximation of the PRK model, for 
effective interference relation estimation, and CPS leverages cyber-physical structures of vehicle traffic flows (particularly, spatiotemporal interference correlation 
as well as macro- and micro-scopic vehicle dynamics) for effective 
use of the gPRK model.
    \ifthenelse{\boolean{ieee10pager}}
    {}
    {
    To support predictable broadcast, CPS controls the interactions between gPRK model adaptations of the links of the same sender to ensure predictable broadcast reliability in the presence of vehicle
    \ifthenelse{\boolean{short}}
    {mobility.}
    {mobility, and we propose a set-cover-based approach to minimum-overhead control signaling.}
    } 
    Through experimental analysis with high-fidelity ns-3 and SUMO simulation, we observe that CPS enables predictable reliability while achieving high throughput and low delay in communication. 
    To the best of our knowledge, CPS is the first field-deployable method that ensures predictable interference control and thus reliability in inter-vehicle communications.
\end{abstract}

\section{Introduction} \label{sec:intro}

Transcending the traditional paradigm of single-vehicle-oriented safety and efficiency control, next-generation vehicles are expected to cooperate with one another and with transportation infrastructures to ensure safety, maximize fuel economy, and minimize emission as well as congestion \cite{Zhang:V2X-survey}.
    One basis for this vision of networked vehicle control (e.g., active safety and fuel economy control \cite{Zhang:V2X-survey}) is wireless communication between close-by vehicles. Critical to the optimality and safety of networked vehicle control, inter-vehicle communication is required to be predictably reliable according to the requirement of vehicle control \cite{Wang:platoonPDR}.
    Given the different impact that communication reliability, delay, and throughput have on networked vehicle control \cite{Wang:platoonPDR,Wang:platoonDelay} and the inherent tradeoff between communication reliability, delay, and throughput \cite{Haenggi:TDR-tradeoff,PRK}, the optimal operation of networked vehicle systems also requires controlling the tradeoffs between communication reliability, delay, and throughput, for which controlling communication reliability in a predictable manner is a basis \cite{control-comm-netRT,PRK}.

Despite extensive research in inter-vehicle wireless networking and pilot field-deployments of IEEE 802.11p-based networks, there still lack solutions for ensuring predictable inter-vehicle communication reliability. Inheriting the basic design principles of WiFi such as CSMA-based channel access control, for instance, existing 802.11p-based solutions may not even be able to ensure a communication reliability of 30\% 
 \cite{Ma:DSRC-reliability-analysis,PRKS}.
    One major reason for the unpredictability and low reliability in existing inter-vehicle wireless networking solutions is the lack of predictable interference control. Thus scheduling data transmissions to control interference in a predictable manner is a basic element of inter-vehicle networking. 

 Given the pervasiveness of vehicles, networks of vehicles tend to be of large scale even though most networked vehicle control only involve communications between close-by vehicles \cite{Zhang:V2X-survey}.
    In the meantime, vehicle mobility introduces dynamics in network topology which, together with uncertainties in wireless communication, introduces complex dynamics and uncertainties in inter-vehicle communication. For agile adaptation to uncertainties and for avoiding information inconsistency in centralized scheduling in large-scale V2V networks, distributed scheduling becomes desirable for interference control in inter-vehicle communications.
    Because wireless signals propagate far away in space and signals from different vehicles add to one other, however, inter-vehicle interference relations tend to be non-local and combinatorial, and predictable interference control tends to require coordination between transmitters far away from one another, which is challenging in highly-dynamic, large-scale V2V networks.

For predictable interference control in distributed scheduling, Zhang et$.$ al \cite{PRK} have identified the physical-ratio-K (PRK) interference model that transforms non-local interference control problems into local control problems which only require explicit coordination between close-by transmitters in scheduling. Based on the PRK model, 
Zhang et$.$ al \cite{PRKS} have also proposed the PRK-based scheduling protocol PRKS which ensures predictable communication reliability in networks of no or low node mobility.
	Not targeting V2V networks, however, PRKS does not address the challenges of \emph{vehicle mobility} to PRK-based scheduling, and it is not applicable to inter-vehicle communications.
	In V2V networks, vehicle mobility makes network topology and inter-vehicle channel properties highly dynamic, which in turn makes interference relations between vehicles highly dynamic, especially for vehicles on different roads or in opposite driving directions of a same road. The highly dynamic nature of inter-vehicle interference relations challenges the precise identification of interference relations in terms of both interference relation estimation and the signaling of interference relations.
    \ifthenelse{\boolean{ieee10pager}}
    {}
	{Additionally, PRKS focuses on predictable unicast reliability without considering predictable reliability in \emph{broadcast} which is a fundamental primitive in inter-vehicle communications \cite{Zhang:V2X-survey}.}
     Thus the open question is \emph{whether it is feasible and how to apply PRK-based scheduling in V2V networks so that the interference between concurrently transmitting vehicles is controlled in a predictable manner to ensure the required inter-vehicle communication reliability.}

In this paper, we give a constructive, positive answer to the question by developing the Cyber-Physical Scheduling (CPS) framework that leverages cyber-physical structures of V2V networks to address the challenges of vehicle
\ifthenelse{\boolean{ieee10pager}}
{mobility,}
{mobility and broadcast,}
and we make the following contributions:
\begin{itemize}
\item For effective control signaling of fast-varying interference relations and by leveraging the physical locations of vehicles, we propose a geometric approximation of the PRK interference model, denoted as the gPRK model. The gPRK model enables vehicles to learn their mutual interference relations in the presence of vehicle mobility and without requiring significant control signaling bandwidth.

\item For accurate identification of interference relations in the presence of vehicle mobility, we propose to leverage cyber-physical structures of vehicle traffic flows (particularly, spatiotemporal interference correlation as well as macro- and micro-scopic vehicle dynamics) for agile instantiation and effective use of the gPRK model in scheduling.

\ifthenelse{\boolean{ieee10pager}}
{}
{
\item For ensuring predictable broadcast reliability in the presence of network dynamics such as vehicle mobility, we address the interactions between gPRK model adaptations of the links of the same broadcast
    \ifthenelse{\boolean{short}}
    {sender.}
    {sender, and we propose a set-cover-based approach to minimizing the overhead of control signaling.}
}

\item We propose the distributed Cyber-Physical Scheduling (CPS) framework that integrates the above 
interference modeling mechanisms in scheduling inter-vehicle communications. We implement CPS in ns-3 \cite{ns-3}, and we experimentally analyze CPS through integrated, high-fidelity simulation of wireless networks and vehicle dynamics using ns-3 and SUMO \cite{SUMO} respectively. 
    We validate 
    that CPS ensures predictable reliability while achieving high throughput and low delay in inter-vehicle communication, thus providing a wireless networking foundation for networked vehicle control.
\end{itemize}
Note that, even though concepts such as vehicle location, vehicle mobility, and wireless channel correlation have been used in various forms in existing protocols, CPS is the first approach that is able to effectively leverage vehicle location information, spatiotemporal interference correlation, vehicle dynamics, and the gPRK model to ensure predictable interference control and thus predictable reliability in inter-vehicle communication, which is non-trivial and is also a critical enabler of the vision of networked vehicle control.

The rest of the paper is organized as follows. In Section~\ref{sec:preliminaries}, we present the system model and problem specification, and we review the PRK interference model \cite{PRK} and PRKS scheduling protocol \cite{PRKS}. We give an overview of the CPS framework in Section~\ref{sec:designOverview}, and then we present our approaches to
\ifthenelse{\boolean{ieee10pager}}
{addressing vehicle mobility in Sections~\ref{sec:addressMobility}.}
{addressing vehicle mobility and broadcast in Sections~\ref{sec:addressMobility} and \ref{sec:addressBroadcast} respectively.}
We experimentally analyze CPS in Section~\ref{sec:eval}, and we discuss related work in Section~\ref{sec:relatedWork}. We make concluding remarks in Section~\ref{sec:concludingRemarks}.

\section{Preliminaries}  \label{sec:preliminaries}

\subHeadingS {System model \& problem specification.}
In inter-vehicle wireless communication networks, referred to as \emph{V2V networks} hereafter, a fundamental communication primitive is single-hop broadcast via which a vehicle shares its states (e.g., location and speed) with close-by vehicles within a certain distance (e.g., 150 meters) \cite{Zhang:V2X-survey}.
Given the significance of single-hop broadcast (e.g., for real-time networked vehicle control \cite{Zhang:V2X-survey}) and for conciseness of presentation, our discussion in this paper focuses on single-hop broadcast, but the proposed methodology for scheduling inter-vehicle broadcasts applies to the scheduling of inter-vehicle single-hop unicast.
Even though we only consider single-hop broadcasts by individual vehicles, we do consider real-world settings where the individual vehicles are widely distributed in space and may well be beyond the broadcast range of many other vehicles.

With the above V2V network setup, we study the online slot-scheduling problem \cite{iOrder} where, given a set of vehicles on the road at any time instant, a maximal subset of the vehicles need to be scheduled in a distributed manner to transmit concurrently while ensuring that the mean packet delivery reliability (PDR) from every transmitting vehicle $S$ to each of its broadcast receivers $R$ is no less than an application-required PDR $T_{S,R}$. Note that a vehicle $R$ is a broadcast receiver of a transmitting vehicle $S$ if the Euclidean distance between $S$ and $R$, denoted by $D(S,R)$, is no more than the communication range of $S$, denoted by $D_S$.
	Focusing on predictable co-channel interference control in broadcast scheduling, we assume that all vehicles share a single communication channel 
and that the broadcast transmission power is fixed for each vehicle even though different vehicles may use different transmission powers; multi-channel scheduling and broadcast power control are relegated as future research.

\subHeading{PRK interference model \& PRKS.}
Despite decades of research in interference-oriented channel access scheduling, most existing literature are either based on the protocol interference model or the physical interference model, neither of which is a good foundation for distributed interference control in the presence of uncertainties \cite{PRK,PRKS}.
    The protocol model is local and suitable for distributed protocol design, but it is inaccurate and does not ensure reliable data delivery \cite{Samir:interference-model}. 
    The physical model has high-fidelity, but it is non-local and combinatorial and thus not suitable for distributed protocol design in dynamic, uncertain network settings \cite{PRK,PRKS}.
\ifthenelse{\boolean{short}}
{}
{
 Even though many centralized TDMA scheduling algorithms have been proposed based on the physical model \cite{Doug:schFramework,geometric-SINR}, distributed physical-model-based scheduling still has various drawbacks such that it cannot ensure predictable communication reliability in practice \cite{PRKS}.
Distributed scheduling algorithms using general pairwise interference models have also been proposed \cite{Srikant:Q-CSMA}. 
Theoretical in nature, however, these algorithms did not address the important question of how to identify the interference set of each link, and their implementation usually assumes a model similar to the protocol model \cite{Srikant:Q-CSMA}.

}
To bridge the gap between the existing interference models and the design of distributed, field-deployable scheduling protocols with predictable communication reliability, Zhang et$.$ al \cite{PRK} have identified \emph{the physical-ratio-K (PRK) interference model} that integrates the protocol model's locality with the physical model's high-fidelity.
In the PRK model, a node $C'$ is regarded as not interfering and thus can transmit concurrently with the transmission from another node $S$ to its receiver $R$  if and only if
$P(C', R) < \frac{P(S, R)}{K_{S, R, T_{S,R}}}$,
where $P(C', R)$ and $P(S, R)$ is the average strength of signals reaching $R$ from $C'$ and $S$ respectively, $K_{S, R, T_{S,R}}$ is the minimum real number 
chosen such that, in the presence of cumulative interference from all concurrent transmitters, the probability for $R$ to successfully receive packets from $S$ is no less than the minimum link relia-
\ifthenelse{\boolean{short} \AND \NOT \boolean{ieee10pager}}
{\begin{wrapfigure}{r}{.4\linewidth}
\centering
\includegraphics[width=.96\linewidth]{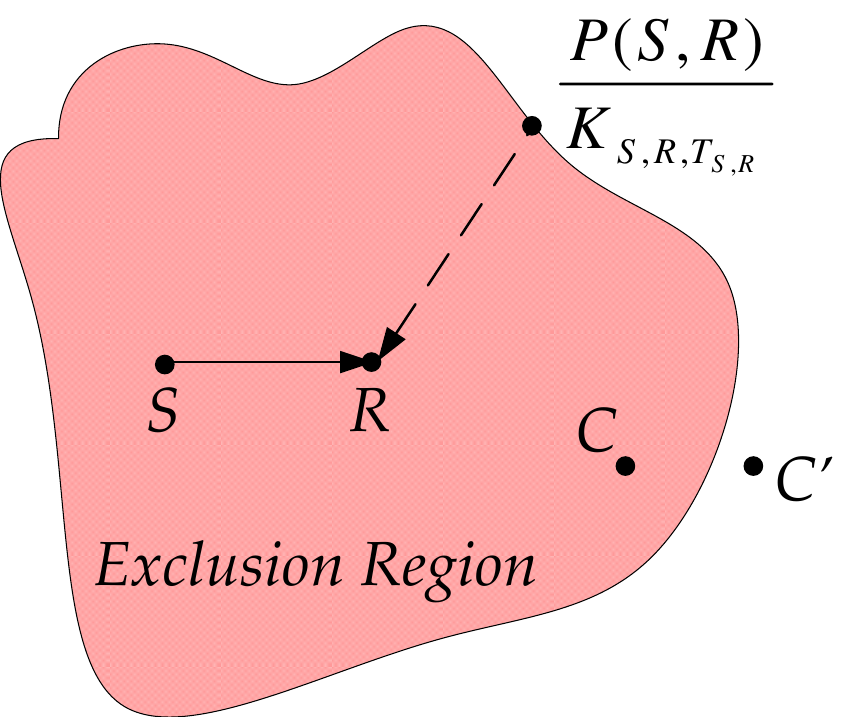}
\caption{PRK interference model} \label{fig:PRK-model}
\vspace*{-0.1in}
\end{wrapfigure}
}{}
bility $T_{S,R}$ required by applications.
\ifthenelse{\boolean{ieee10pager}}
{}
{
As shown in Figure~\ref{fig:PRK-model}, the PRK model defines, for each link $\langle S, R \rangle$, an exclusion region (ER) $\mathbb{E}_{S,R,T_{S,R}}$ around the receiver $R$ such that a node $C$ is in the ER (i.e., $C \in \mathbb{E}_{S,R,T_{S,R}}$) if and only if $P(C, R) \ge \frac{P(S, R)}{K_{S,R,T_{S,R}}}$. Every node $C \in \mathbb{E}_{S,R,T_{S,R}}$ is regarded as interfering and thus shall not transmit concurrently with the transmission from $S$ to $R$. 
\ifthenelse{\boolean{short}}
{}
{
\begin{wrapfigure}{r}{.4\linewidth}
\centering
\includegraphics[width=.96\linewidth]{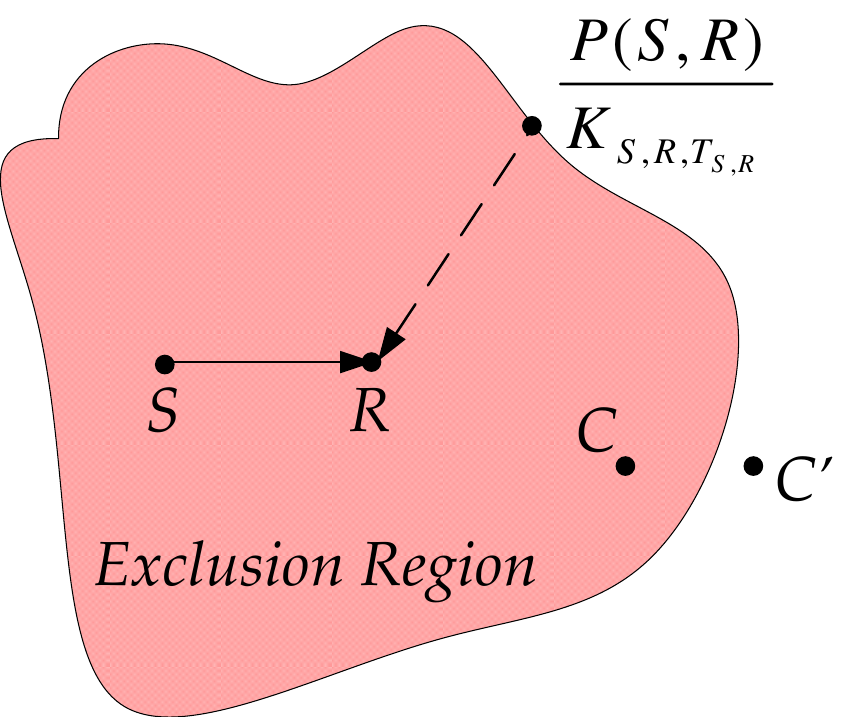}
\caption{PRK interference model} \label{fig:PRK-model}
\vspace*{-0.1in}
\end{wrapfigure}
}
} 

For predictable interference control, 
the parameter $K_{S,R,T_{S,R}}$ of the PRK model needs to be instantiated for every link $\langle S, R \rangle$ according to in-situ, potentially unpredictable network and environmental conditions (e.g., data traffic load and wireless signal power attenuation).
To this end, Zhang et$.$ al \cite{PRKS} have formulated the PRK model instantiation problem as a regulation control problem \cite{book:feedbackControl} where the ``plant'' is the link $\langle S, R \rangle$, the ``reference input'' is the required link reliability $T_{S,R}$, the ``output'' is the actual link reliability $Y_{S,R}$ from $S$ to $R$, the ``control input'' is the PRK model parameter $K_{S,R,T_{S,R}}$, and the objective of the regulation control is to adjust the control input so that the plant output is no less than the reference input.
	Then 
control theory can be used to derive the controller for instantiating the PRK model parameter \cite{PRKS}.
	For every link $\langle S, R \rangle$, using its instantiated PRK model parameter $K_{S,R,T_{S,R}}$ and the local \emph{signal maps} that contain average signal power between $S$, $R$, and every other close-by node $C$ that may interfere with the transmission from $S$ to $R$, link $\langle S, R \rangle$ and every close-by node $C$ become aware of their mutual interference relations.
With precise awareness of mutual interference relations with close-by nodes/links, nodes schedule data transmissions in a 
TDMA fashion using the distributed optimal-node-activation-multiple-access (ONAMA) algorithm \cite{ONAMA}, 
and the resulting PRK-based scheduling protocol is denoted as PRKS \cite{PRKS}.
\ifthenelse{\boolean{short}}
{Through extensive measurement study in the high-fidelity Indriya \cite{Indriya} and NetEye \cite{NetEye} wireless network testbeds, Zhang et$.$ al \cite{PRKS} observe that PRKS enables predictable interference control while achieving high channel spatial reuse. Accordingly, PRKS enables predictable link reliability, high network throughput, and low communication delay \cite{PRKS}.
}
{

Through extensive measurement study in the high-fidelity Indriya \cite{Indriya} and NetEye \cite{NetEye} wireless network testbeds, Zhang et$.$ al \cite{PRKS} observe the following:
	1) The distributed controllers for PRK model instantiation 
enable network-wide convergence to a state where the desired 
link reliabilities are ensured; 
	2) With local, distributed coordination alone, PRKS achieves a channel spatial reuse very close to (e.g., more than 90\%) what is enabled by the state-of-the-art centralized physical-model-based scheduler iOrder \cite{iOrder} while ensuring the required 
link reliability; 
	3) Unlike existing protocol-model- or physical-model-based scheduling protocols where link reliability is unpredictable and the ratio of links whose reliability meets application requirements can be as low as 0\%, PRKS enables predictably high link reliability (e.g., 95\%) for all the links in different network and environmental conditions without a priori knowledge of these conditions; 
        4) By ensuring the required 
link reliability in scheduling, PRKS also enables a lower communication latency and a higher network throughput than existing scheduling protocols. 
}

\section{Overview of CPS}  \label{sec:designOverview}


A major challenge in applying PRK-based scheduling to V2V networks is vehicle mobility.
    Vehicle mobility makes inter-vehicle wireless channels highly dynamic, thus, as we will analyze in Section~\ref{subsec:gPRK}, it would be too costly or even infeasible for vehicles to maintain accurate signal maps that store reception power of data signals between close-by vehicles, thus making the PRK interference model and the PRKS scheduling protocol not applicable to V2V networks. To address this challenge, we observe that the physical vehicle locations are readily available in V2V networks through GPS and/or other mechanisms such as simultaneous-localization-and-mapping (SLAM). Accordingly, we propose the gPRK interference model as a geometric approximation of the PRK model, so that the gPRK model enables lightweight approaches for vehicles to detect their mutual interference relations using vehicle locations instead of signal maps.
Vehicle mobility also makes vehicle locations and thus inter-vehicle interference relations highly dynamic. For enabling vehicles to accurately track their mutual interference relations, we propose to leverage spatiotemporal interference correlation and macroscopic vehicle dynamics to quickly instantiate the gPRK model parameters of newly-established and transient links, and to leverage well-understood microscopic vehicle dynamics to 
track vehicle locations.

\ifthenelse{\boolean{ieee10pager}}
{}
{
Another challenge in V2V networks is the need to support predictably reliable inter-vehicle broadcast. To ensure the communication reliability from a sender vehicle $S$ to each of its broadcast receiver vehicles, we define the sender exclusion region (a$.$k$.$a$.$ sender ER) of $S$ as the union of the exclusion region (ER) around each receiver of $S$, and a vehicle in the sender ER of $S$ is regarded as an interferer of $S$ and shall not transmit concurrently with $S$.
    \ifthenelse{\boolean{short}}
    {}
    {For light-weight, effective sharing of sender ERs between vehicles, we propose a set-cover-based approach to minimizing the overhead of control signaling.}
    For broadcast, the exclusion regions (ERs) around the receivers of the same sender overlap with one another, and this makes the gPRK model adaptation of the involved links interact with one another. We propose gPRK model adaptation rules that explicitly address this interaction to ensure predictable broadcast reliability in the presence of vehicle mobility.
} 

\begin{figure}[!htbp]
  \vspace*{-0.1in}
  \centering
  \includegraphics[width=0.8\linewidth]{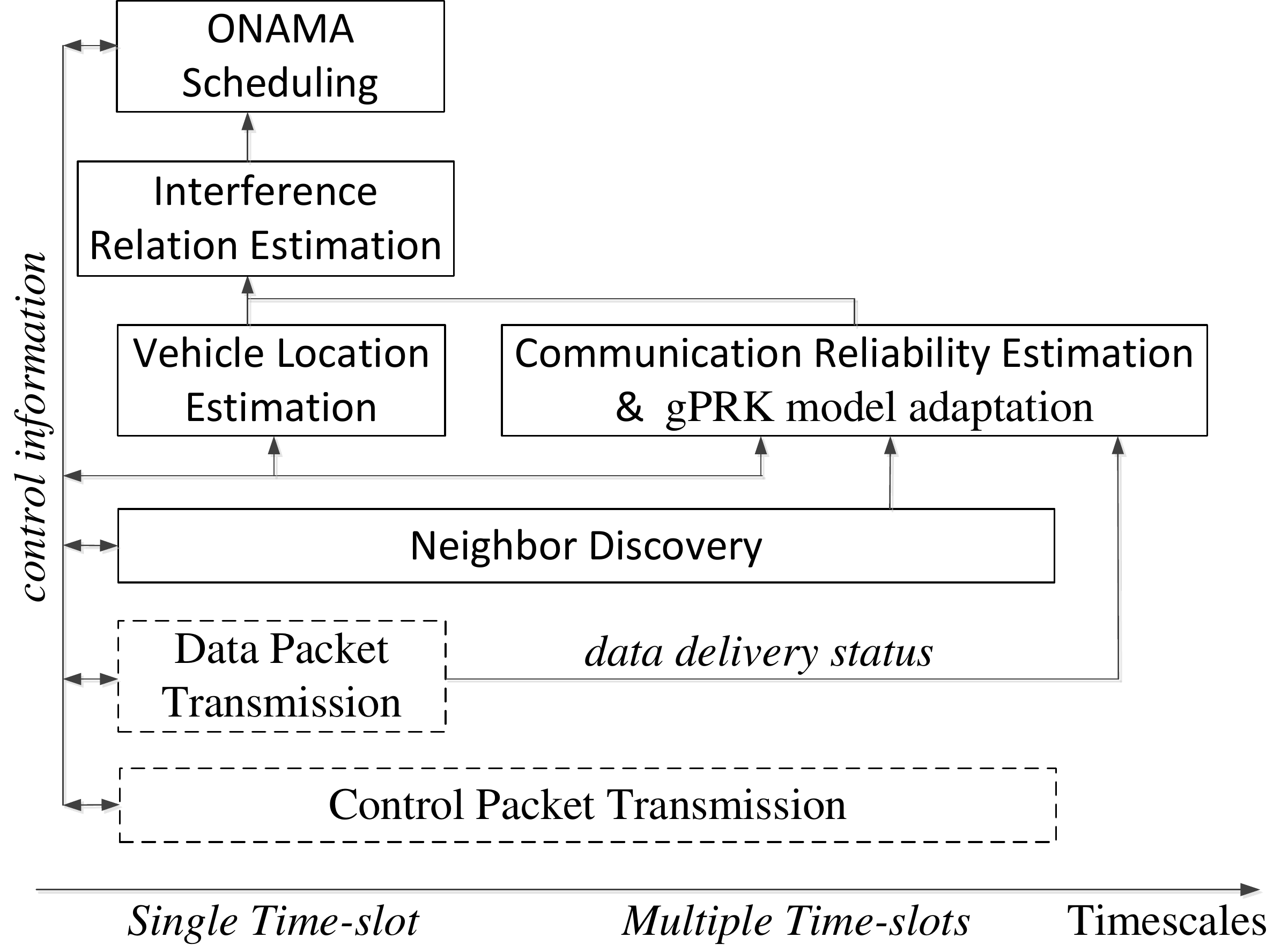}
  \caption{
  Cyber-Physical Scheduling (CPS) framework}   \label{fig:CPS-framework}
  \vspace*{-0.1in}
\end{figure}

Using the above methods of leveraging cyber-physical structures of V2V networks (particularly, spatiotemporal interference correlation, correlated ER adaptation, physical vehicle location, as well as macro- and micro-scopic vehicle dynamics) to address vehicle
\ifthenelse{\boolean{ieee10pager}}
{mobility,}
{mobility and broadcast,}
vehicles can identify their mutual interference relations in an agile, distributed manner. Based on the mutual interference relations among vehicles, inter-vehicle communications can be scheduled in a TDMA manner similar to that in PRKS \cite{PRKS}.
    To realize the above methods, we propose the Cyber-Physical Scheduling (CPS) framework for inter-vehicle communications as shown in Figure~\ref{fig:CPS-framework}.
In this framework, time is divided into time slots, and, as in PRKS \cite{PRKS}, the transmissions of control signaling packets (e.g., those containing gPRK model parameters, vehicle locations, and/or sender ERs) and data packets are separated in frequency or in time so that there is no interference between control packet transmission and data packet transmission.
     Through the exchange of control signaling packets, close-by vehicles discover one another and initialize the gPRK model parameters for the corresponding links.
    Based on feedback on the status (i.e., success or failure) of data packet transmissions, in-situ communication reliabilities are estimated and then gPRK model parameters are adapted on the fly. Together with estimated locations of close-by vehicles, the in-situ gPRK model parameters enable vehicles to detect their mutual interference relations. Based on in-situ interference relations, a maximal set of mutually non-interfering vehicles are scheduled to transmit their data packets at each time slot according to the distributed TDMA algorithm ONAMA 
    \cite{ONAMA}.

From each vehicle's perspective, immediately after it starts, it quickly discovers close-by vehicles, initializes related gPRK model parameters, and detects mutual interference relations with close-by vehicles; then, in parallel with data transmissions and using feedback on data transmission status (i.e., success or failure), the vehicle adapts its gPRK model parameters, and, with adaptive estimation of the locations of close-by vehicles, the vehicle adapts data transmission schedules according to in-situ interference relations with close-by vehicles.
    Figure~\ref{fig:CPS-framework} shows the timescales of different protocol actions in CPS.
    When a vehicle starts, it quickly performs neighbor-discovery at every time slot for a short period (e.g., 2 seconds), and then it maintains neighborhood information at a frequency of regular control packet transmissions (e.g., every 100 time slots).
    Given a vehicle and a link from a sending vehicle, the gPRK model parameter is updated each time a new communication reliability estimation becomes available (e.g., every 1,000 time slots).
    Each vehicle updates its estimation of the locations of close-by vehicles and its interference relations with close-by vehicles every time slot, which enables the ONAMA-based scheduling of non-interfering concurrent transmitters at each time slot.
    In our implementation, we have set the duration of each time slot to be 2.5 milliseconds so that a data packet up to 1,500 bytes can be delivered in a time slot when the radio transmission rate is 6Mbps (i.e., the lowest transmission rate of the current 802.11p standard) and when operations other than the actual data transmission (e.g., composing the data packet) may take up to 0.5 millisecond in a time slot; accordingly, inter-vehicle interference relations and gPRK model parameters are updated every 2.5 milliseconds and about every 2.5 seconds respectively.

With the above overview of the CPS framework, we next elaborate on our approaches to addressing vehicle
\ifthenelse{\boolean{ieee10pager}}
{mobility in Section~\ref{sec:addressMobility}.}
{mobility and broadcast in Sections~\ref{sec:addressMobility} and \ref{sec:addressBroadcast} respectively.}
For conciseness of presentation, our discussion will focus on a sender $S$ and its receiver set $\mathbf{R} = \{R: R \neq S \land D(S,R) < D_S\}$ unless mentioned otherwise.

\section{Addressing Vehicle Mobility}  \label{sec:addressMobility}

\ifthenelse{\boolean{ieee10pager}}
{}
{
Vehicle mobility makes inter-vehicle interference relations highly dynamic, and this challenges 
both interference relation estimation and the signaling of interference relations. In what follows, we present our approaches that leverage cyber-physical structures of V2V networks to address the challenges.
}


\subsection{
Geometric Approximation of PRK Model} \label{subsec:gPRK}

\subHeadingS{Challenge of using PRK model in V2V networks.}
As discussed in Section~\ref{sec:preliminaries}, the definition of the PRK interference model is based on signal power between close-by nodes. To use the PRK model in data transmission scheduling, nodes need to maintain local signal maps so that interfering nodes and links can be aware of their mutual interference relations. For networks of no or low node mobility which Zhang et al$.$ \cite{PRKS} have considered, the average signal power between nodes does not change at timescales such as seconds, minutes, or even hours. Accordingly, the frequency of signal map update and thus the overhead of signal map maintenance tends to be low for networks of no or low mobility \cite{PRKS}.
    For V2V networks, however, vehicle mobility makes average signal power between close-by vehicles fast-varying in nature, for instance, at the timescales of seconds or less. If we were to apply the PRK interference model to V2V networks, the local signal maps between close-by vehicles 
    would need to be updated frequently. 
    In particular, every vehicle $R$ needs to frequently estimate the in-situ signal power from every other potentially interfering vehicle $C$ to itself; after each estimation, $R$ needs to share the newly estimated signal power $P(C,R)$ with every other potentially interfering vehicle through control signalling packet exchange \cite{PRKS}, which would introduce significant messaging overhead.
\ifthenelse{\boolean{short}}
{
As we derive in \cite{CPS-TR}, for a typical network setting, Figure~\ref{fig:signalMapOverhead}
  \begin{figure}[!htbp]
    \vspace*{-0.1in}
    \centering
    \includegraphics[width=\figWidth]{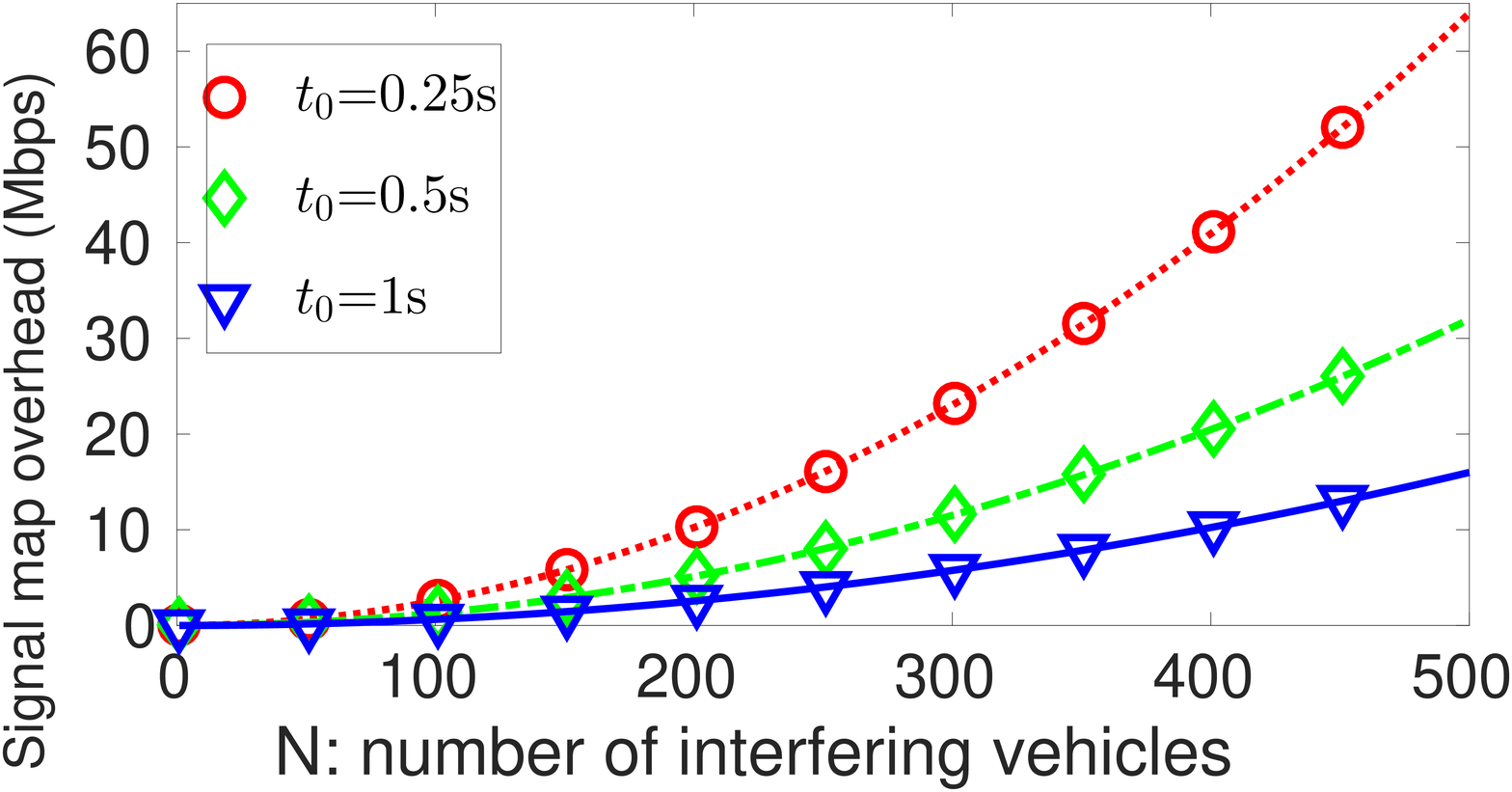}
    \caption{Overhead of signal map maintenance ($t_0$: interval between consecutive signal map updates)} \label{fig:signalMapOverhead}
    \vspace*{-0.1in}
  \end{figure}
shows the significant overhead of signal map maintenance in V2V networks. Considering that the current physical layer of the V2V communication standard IEEE 802.11p can only support a transmission rate of 6Mbps - 27Mbps, that the total bandwidth available to a set of mutually interfering vehicles is no more than the transmission rate, and that $N$ (i.e., the number of interfering vehicles for a vehicle) may well be in the range of hundreds (e.g., in urban settings), Figure~\ref{fig:signalMapOverhead} shows that the signal map maintenance overhead accounts for a significant portion or even exceed the total communication bandwidth of V2V networks. This implies that it is \emph{too costly or even infeasible to maintain accurate signal maps for PRK-based scheduling in V2V networks}.
    Therefore, it is difficult, if not impossible, to directly apply the PRK interference model to data transmission scheduling in V2V networks.
}
{

Assuming there are $N$ close-by vehicles that may interfere with one another, for instance, the signal map would contain, for every vehicle $v_i (i = 1 \ldots N)$, the average signal power from every other vehicle to $v_i$. Since every vehicle $v_i$ can only estimate the average signal power from every other vehicle to itself through received-signal-strength-indicator (RSSI) sampling \cite{PRKS}, it is necessary for every vehicle $v_i$ to share its estimates with every other vehicle in order for every vehicle to establish and maintain its own local signal map about the signal power between close-by vehicles. 
    For instance, a receiver vehicle $R$ can sample and estimate the signal power $P(C, R)$ from another vehicle $C$ to itself, but $R$ has to shared its estimate of $P(C,R)$ with $C$ in order for $C$ to know $P(C,R)$ and thus decide whether itself can interfere with the transmission from a sender vehicle $S$ to $R$ based on the PRK model.
Assuming it takes two bytes to encode the signal power from one vehicle to another and it takes six bytes to encode the ID (e.g., MAC address) of each vehicle, it takes $(6 + 8(N-1))$ bytes for a vehicle $v_i$ to encode the signal power from every other vehicle to itself. Therefore, each update of the signal map takes $N(6 + 8(N-1))$ bytes of information exchange between vehicles. Assuming the signal map is updated every $t_0$ seconds, the signal map maintenance will consume $\frac{8N(6 + 8(N-1))}{t_0}$bps network bandwidth.
    For typical values of $N$ in V2V networks and different update intervals $t_0$, Figure~\ref{fig:signalMapOverhead}
  \begin{figure}[!htbp]
    \vspace*{-0.1in}
    \centering
    \includegraphics[width=\figWidth]{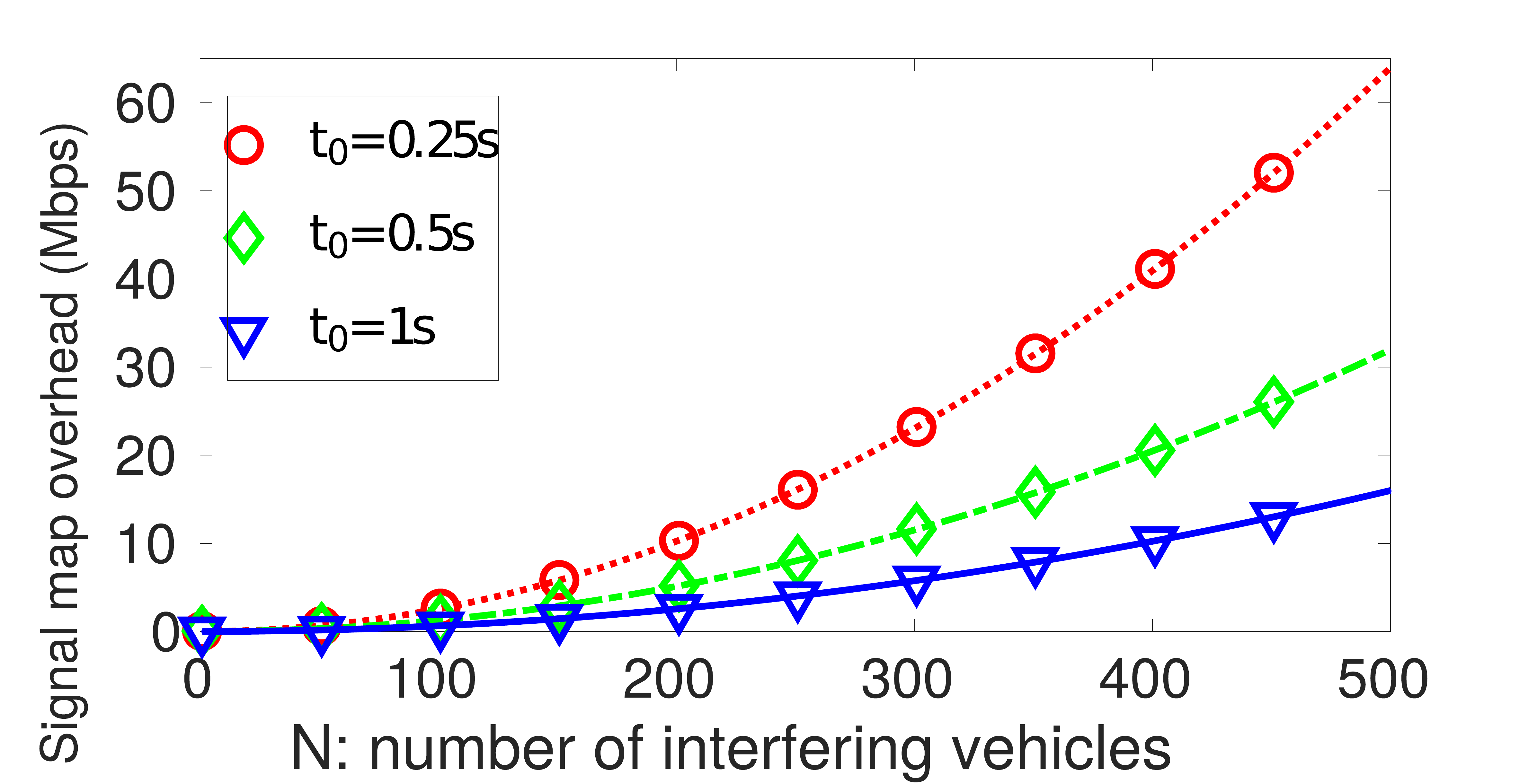}
    \caption{Overhead of signal map maintenance} \label{fig:signalMapOverhead}
    \vspace*{-0.2in}
  \end{figure}
shows the significant overhead of signal map maintenance in V2V networks.
    Considering that the current physical layer of the V2V communication standard IEEE 802.11p can only support a transmission rate of 6Mbps - 27Mbps, that the total bandwidth available to a set of mutually interfering vehicles is no more than the transmission rate, and that $N$ may well be in the range of hundreds (e.g., in urban settings), Figure~\ref{fig:signalMapOverhead} shows that the signal map maintenance overhead accounts for a significant portion or even exceed the total communication bandwidth of V2V networks. This implies that it is \emph{too costly or even infeasible to maintain accurate signal maps for PRK-based scheduling in V2V networks}.
    Therefore, it is difficult, if not impossible, to directly apply the PRK interference model to data transmission scheduling in V2V networks.
}

\subHeading{gPRK interference model.}
In V2V network systems, vehicle locations are important factors for networked vehicle control, and thus they are readily available through GPS and/or other mechanisms such as simultaneous-localization-and-mapping (SLAM). Using vehicle locations, it is easy for vehicles to know the distances among themselves. To avoid the significant overhead (and sometimes infeasibility) of maintaining accurate signal maps in V2V networks and considering the fact that, on average, closer-by vehicles tend to introduce higher interference signal power to one another than farther away vehicles, we propose to leverage the availability of vehicle location information to define a geometric approximation of the PRK interference model, denoted as the \emph{gPRK interference model}.
     In the gPRK model, interference relations among vehicles are defined based on inter-vehicle distance instead of inter-vehicle signal power, and a vehicle $C'$ is regarded as not interfering and thus can transmit concurrently with the transmission from another vehicle $S$ to its receiver $R$  if and only if
\begin{equation}\label{eqn:gPRK-model}
D(C', R) > D(S, R) K_{S, R, T_{S,R}},
\end{equation}
where $D(C', R)$ and $D(S, R)$ is the geometric distance between $C'$ and $R$ and that between $S$ and $R$ respectively, $K_{S, R, T_{S,R}}$ is the minimum real number 
chosen such that, in the presence of cumulative
  \begin{wrapfigure}{r}{.4\linewidth}
    \vspace*{-0.1in}
    \centering
    \includegraphics[width=0.99\linewidth]{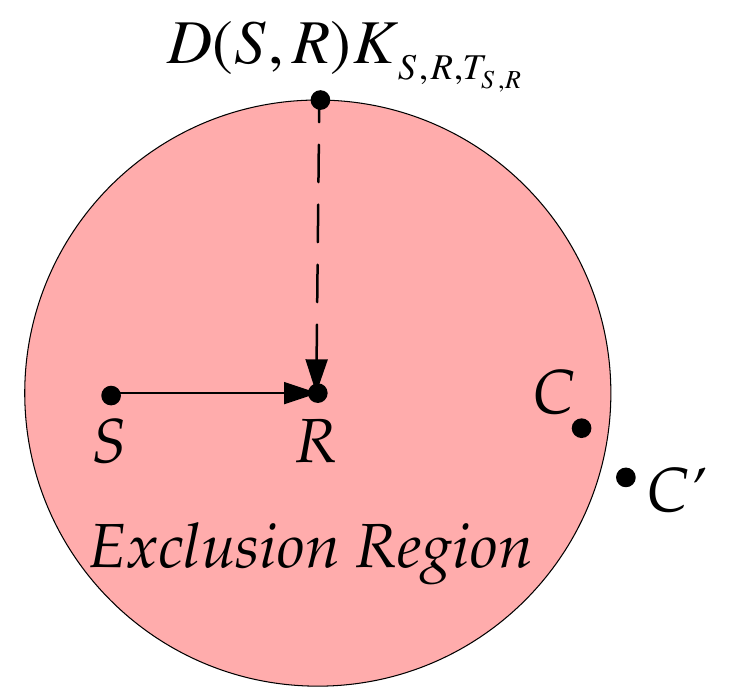}
    \caption{gPRK 
    model} \label{fig:gPRK-model}
    \vspace*{-0.1in}
  \end{wrapfigure}
interference from all concurrent transmitters, the probability for $R$ to successfully receive packets from $S$ is no less than the minimum link reliability $T_{S,R}$ required by applications.
    As shown in Figure~\ref{fig:gPRK-model},
the gPRK model defines, for each link $\langle S, R \rangle$, an exclusion region (ER) $\mathbb{E}_{S,R,T_{S,R}}$ around the receiver $R$ such that a node $C$ is in the region (i.e., $C \in \mathbb{E}_{S,R,T_{S,R}}$) if and only if $D(C, R) \le D(S, R) K_{S,R,T_{S,R}}$.

Similar to the PRK model, the gPRK model is \emph{local} since only local, pairwise interference relations are defined between close-by vehicles. Similar to the PRK model and unlike the protocol interference model which is not adaptive to network and environmental conditions and thus unable to ensure predictable interference control, the gPRK model is suitable for \emph{reliable inter-vehicle communication} since it ensures the application-required link reliability by adapting parameter $K_{\ldots}$ according to in-situ network and environmental conditions and by considering wireless communication properties such as cumulative interference. 
    Unlike the PRK model where the ER around a link may be of an irregular shape due to anisotropic wireless signal propagation \cite{PRK}, the ER around a link in the gPRK model is of the regular shape of a disk. 
\ifthenelse{\boolean{short}}
{
As we elaborate in detail in \cite{CPS-TR}, 
this difference between the gPRK and PRK models becomes insignificant for inter-vehicle
\ifthenelse{\boolean{temp}}
{broadcast for the following reasons: firstly, the exclusion region of a broadcast is the union of the exclusion regions of all the links from the broadcast sender to the individual receivers, as we will discuss in detail 
\ifthenelse{\boolean{ieee10pager}}
{shortly;}
{in Section~\ref{sec:addressBroadcast};}
secondly, a vehicle in the PRK-based (or gPRK-based) ER of a link $l_i$ but not in the gPRK-based (or PRK-based) ER of $l_i$ may still be in the gPRK-based (or PRK-based) ER of another link $l_j$ ($j \ne i$) and thus in the gPRK-based (or PRK-based) exclusion region of the broadcast.
}
{broadcast, and, ???in Section~\ref{sec:eval}, we experimentally show that gPRK-based scheduling enables statistically the same concurrency as PRK-based scheduling while ensuring the application-required reliability.}

With the gPRK model, a vehicle only needs to share its location with potentially interfering vehicles in order for an interfering vehicle to detect their mutual interference relation using the gPRK model parameter $K$, and a vehicle does not need to share with other vehicles the signal power from every other potentially interfering vehicle to itself. As we show in \cite{CPS-TR}, this enables \emph{orders of magnitude reduction in control signaling overhead}, which in turn makes it feasible and efficient to use the gPRK model in real-world V2V networks.
}
{
As we will discuss in Section~\ref{sec:addressBroadcast}, this difference between the gPRK and PRK models becomes insignificant for inter-vehicle
\ifthenelse{\boolean{temp}}
{broadcast. }
{broadcast, and, ???in Section~\ref{sec:eval}, we experimentally show that gPRK-based scheduling enables statistically the same concurrency as PRK-based scheduling while ensuring the application-required reliability.}

With the gPRK model, a vehicle only needs to share its location with potentially interfering vehicles in order for an interfering vehicle to detect their mutual interference relation using the gPRK model parameter $K$, and a vehicle does not need to share with other vehicles the signal power from every other potentially interfering vehicle to itself.
    With seven bytes, a vehicle can encode its longitude and latitude information such that the location information accuracy is 1.11meters. Then, for the case of $N$ mutually-interfering vehicles as discussed earlier and assuming it takes six bytes to encode the ID (e.g., MAC address) of a vehicle, it takes $13N$ bytes of information exchange between vehicles in order for the N vehicles to be mutually aware of one another's location. Assuming that the location update frequency is the same as that of signal map update in PRK-based scheduling, using the gPRK model instead of the PRK model would reduce the control signaling overhead by a factor of $\frac{8N(6+8(N-1))}{13N} = \frac{48}{13} + \frac{64}{13}(N-1)$. Using location prediction via microscopic vehicle dynamics models as we will discuss in Section~\ref{sec:addressMobility}, the update frequency of vehicle locations can be lower than that of signal map, thus enabling more reduction in control overhead.
For highly reliable inter-vehicle communication in large-scale V2V networks, $N$ tends to be large and in the range of hundreds. 
Thus the use of the gPRK model instead of the PRK model enables \emph{orders of magnitude reduction in control signaling overhead}, which in turn makes it feasible and efficient to use the gPRK model in real-world V2V networks.
} 

\subHeadingS{gPRK model adaptation.}
Similar to the PRK model, the parameter $K_{S,R,T_{S,R}}$ of the gPRK model needs to be instantiated for every link $\langle S, R \rangle$ according to in-situ 
network and environmental conditions such as vehicle spatial distribution and wireless signal power attenuation. To this end, we use the control-theoretic approach of Zhang et al$.$ \cite{PRKS} that, upon a feedback on the link reliability of $\langle S, R \rangle$ at a time instant $t$, denoted by $Y_{S,R}(t)$, computes the change of cumulative interference power at the receiver $R$, denoted by $\Delta I_R(t)$, that the change of $K_{S,R,T_{S,R}}$ at time $t$ needs to introduce to make $Y_{S,R}(t+1) = T_{S,R}$ at the next time instant $t+1$.\footnote{In protocol implementation, the actual time interval between time instants $t$ and $t+1$ is the time interval for $R$ to compute its $(t+1)$-th sample of communication reliability along $\langle S, R \rangle$.}
In particular, letting $y(t) =  c y(t-1) + (1-c)Y_{S,R}(t) \ (0 \le c < 1)$, $\Delta I_R(t)$ is computed as follows \cite{PRKS}:
\begin{equation} \label{eqn:Delta-I_R}
\Delta I_R(t) =  \frac{(1+c)y(t) - cy(t-1)-T_{S,R}}{(1-c)a(t)} - \mu_U(t),
\end{equation}
where $a(t) = \frac{T_{S,R} - Y_{S,R}(t)}{f^{-1}(T_{S,R}) - f^{-1}(Y_{S,R}(t))}$, 
$f(.)$ is the radio model function that defines the relation between link reliability $Y_{S,R}(t)$ and the signal-to-interference-plus-noise-ratio (SINR) at the receiver $R$ at time $t$, and $\mu_U(t)$ denotes the mean change of the cumulative interference power that vehicles not in $\mathbb{E}_{S,R,T_{S,R}}(t) \cup \mathbb{E}_{S,R,T_{S,R}}(t+1)$ introduce to the receiver $R$ from time $t$ to $t+1$.

Since the receiver $R$ can locally measure or estimate $y(t), y(t-1), a(t)$, and $\mu_U(t)$ \cite{PRKS}, $R$ can locally compute $\Delta I_R(t)$. After computing $\Delta I_R(t)$ at time $t$, $R$ needs to compute $K_{S,R,T_{S,R}}(t+1)$ so
\ifthenelse{\boolean{short}}
{}
{that
\begin{equation} \label{eqn:K-adapt}
\begin{cases}
K_{S,R,T_{S,R}}(t+1) = K_{S,R,T_{S,R}}(t),        & \textrm{if} \ \Delta I_R(t) = 0 \\
K_{S,R,T_{S,R}}(t+1) > K_{S,R,T_{S,R}}(t),   & \textrm{if} \ \Delta I_R(t) < 0 \\
K_{S,R,T_{S,R}}(t+1) < K_{S,R,T_{S,R}}(t),        & \textrm{if} \ \Delta I_R(t) > 0 \\
\end{cases}
\end{equation}
and
}that the expected link reliability is no less than $T_{S,R}$ when the PRK model parameter is $K_{S,R,T_{S,R}}(t+1)$,\footnote{Due to the discrete nature of node distribution, the resulting link reliability may be slightly higher than the required reliability $T_{S,R}$ instead of being exactly $T_{S,R}$. } and, when the PRK model parameter is $\min \{K_{S,R,T_{S,R}}(t), K_{S,R,T_{S,R}}(t+1)\}$, the expected interference introduced to $R$ by the nodes in either $\mathbb{E}_{S,R,T_{S,R}}(t)$ or $\mathbb{E}_{S,R,T_{S,R}}(t+1)$ but not in both is as close to $|\Delta I_R(t)|$ as possible to ensure as high channel spatial reuse as possible. 

To realize the above design, we define, for each node $C$ that may be included in the exclusion region of $R$ during network operation, the expected interference $I(C, R, t)$ that $C$ introduces to $R$ when $C$ is not in the exclusion region of $R$. Then $I(C, R, t) = \beta_C(t) P(C, R, t)$, where $\beta_C(t)$ is the probability for $C$ to transmit data packets at time $t$, $P(C, R, t)$ is the power strength of the data signals reaching $R$ from $C$, and $R$ can estimate $P(C, R, t)$ and $\beta_C(t)$ by passively monitoring the control signaling packets transmitted by $C$ without introducing additional control signal packets \cite{PRKS}.
	Considering the discrete nature of node distribution in space and the requirement on satisfying the minimum link reliability $T_{S,R}$, we propose the following rules for computing $K_{S,R,T_{S,R}}(t+1)$:
\begin{itemize}
\item \ruleHeading{ER0}
If $\Delta I_R(t) = 0$, let $K_{S,R,T_{S,R}}(t+1) = K_{S,R,T_{S,R}}(t)$.

\item \ruleHeading{ER1}
If $\Delta I_R(t) < 0$ (i.e., need to expand the exclusion region), let $\mathbb{E}_{S,R,T_{S,R}}(t+1) = \mathbb{E}_{S,R,T_{S,R}}(t)$, then keep adding nodes not already in $\mathbb{E}_{S,R,T_{S,R}}(t+1)$, in the non-decreasing order of their distance to $R$, into $\mathbb{E}_{S,R,T_{S,R}}(t+1)$ until the node $B$ such that adding $B$ into $\mathbb{E}_{S,R,T_{S,R}}(t+1)$ makes $\sum_{C \in \mathbb{E}_{S,R,T_{S,R}}(t+1) \setminus \mathbb{E}_{S,R,T_{S,R}}(t)} I(C, R, t) \ge |\Delta I_R(t)|$ for the first time. Then let $K_{S,R,T_{S,R}}(t+1) = \frac{D(B,R,t)}{D(S,R,t)}$, where $D(B,R,t)$ and $D(S,R,t)$ denote the distance from $B$ and $S$ to $R$ at time $t$ respectively.

\item \ruleHeading{ER2}
\sloppy{
If $\Delta I_R(t) > 0$ (i.e., need to shrink the exclusion region), let $\mathbb{E}_{S,R,T_{S,R}}(t+1) = \mathbb{E}_{S,R,T_{S,R}}(t)$, then keep removing nodes out of $\mathbb{E}_{S,R,T_{S,R}}(t+1)$, in the non-increasing order of their distance to $R$, until the node $B$ such that removing any more node after removing $B$ makes $\sum_{C \in \mathbb{E}_{S,R,T_{S,R}}(t) \setminus \mathbb{E}_{S,R,T_{S,R}}(t+1)} I(C, R, t) > \Delta I_R(t)$ for the first time. Then let $K_{S,R,T_{S,R}}(t+1) = \frac{D(B,R,t)}{D(S,R,t)}$.
}
\end{itemize}
\ifthenelse{\boolean{short}}
{For convenience, we call the above rules the \emph{gPRK-model-adaptation} rules.  (An example of gPRK model adaptation can be found in \cite{CPS-TR}.) }
{
Figure~\ref{fig:ER-change} demonstrates the above idea for cases when $\Delta I_R(t) \ne 0$. For convenience, we call the above rules the \emph{gPRK-model-adaptation} rules.
\begin{figure}[!htbp]
\centering
\includegraphics[width=.9\linewidth]{ER-change}
\caption{Computing $K_{S,R,T_{S,R}}(t+1)$} \label{fig:ER-change}
\end{figure}
}
As link $\langle S, R \rangle$ updates its parameter $K_{S,R,T_{S,R}}$, the parameter is shared with nodes within $\mathbb{E}_{S,R,T_{S,R}}$ through control signaling as discussed in Section~\ref{sec:designOverview}, which enables nodes to be aware of their mutual interference relations and thus to schedule transmissions with predictable interference control.

\ifthenelse{\boolean{ieee10pager}}
{
\subHeading{Supporting predictable broadcast.}
A fundamental communication primitive in V2V networks is single-hop broadcast via which a vehicle shares its state (e.g., location and speed) with close-by vehicles within a certain distance \cite{Zhang:V2X-survey}. Reliable broadcast is a well-known challenge because, even though acknowledgments from receivers are required for many reliability-enhancement mechanisms such as ACK-/negative-ACK-based retransmission of lost packets and RTS-CTS-based collision avoidance in medium access control, it is difficult for a sender to reliably and efficiently get an acknowledgment from every receiver, especially when the number of receivers is large in V2V networks (e.g., up to hundreds).

To address the challenge, we observe that, to ensure a minimum broadcast reliability $T_S$ for a sender $S$, we need to make sure that the communication reliability along the link from  $S$ to every one of its receiver $R_i \in \mathbf{R}$ is at least $T_S$. This fact enables us to define, for a broadcast sender $S$, a \emph{receiver exclusion region (ER)} $\mathbb{E}_{S,R_i,T_S}$ for every receiver $R_i \in \mathbf{R}$ based on the gPRK model. Accordingly, we define the \emph{sender ER} for $S$, denoted by $\mathbb{E}_{S,T_S}$,  as the union of
its corresponding receiver ERs; that is, $\mathbb{E}_{S,T_S} = \cup_{R_i \in \mathbf{R}} \mathbb{E}_{S,R_i,T_S}$. 
    Based on the definition of the sender ER, the broadcast reliability of $T_S$ is ensured as long as no node in $\mathbb{E}_{S,T_S}$ transmits concurrently with sender $S$.

}
{}

\subsection{gPRK Modeling with Vehicle Mobility}  \label{subsec:gPRK-vehicle-mobility}

Vehicle mobility makes network topology and interference relations highly dynamic (especially for vehicles on different roads or in opposite driving directions of a same road), and this challenges the instantiation and use of the gPRK model in V2V networks. In what follows, we elaborate on our design that addresses the challenges by effectively leveraging cyber-physical structures of V2V networks, particularly, the spatiotemporal interference correlation as well as 
 macro- and micro-scopic vehicle dynamics.

\begin{wrapfigure}{r}{.5\linewidth}
    \vspace*{-0.1in}
    \centering
    \includegraphics[width=\linewidth]{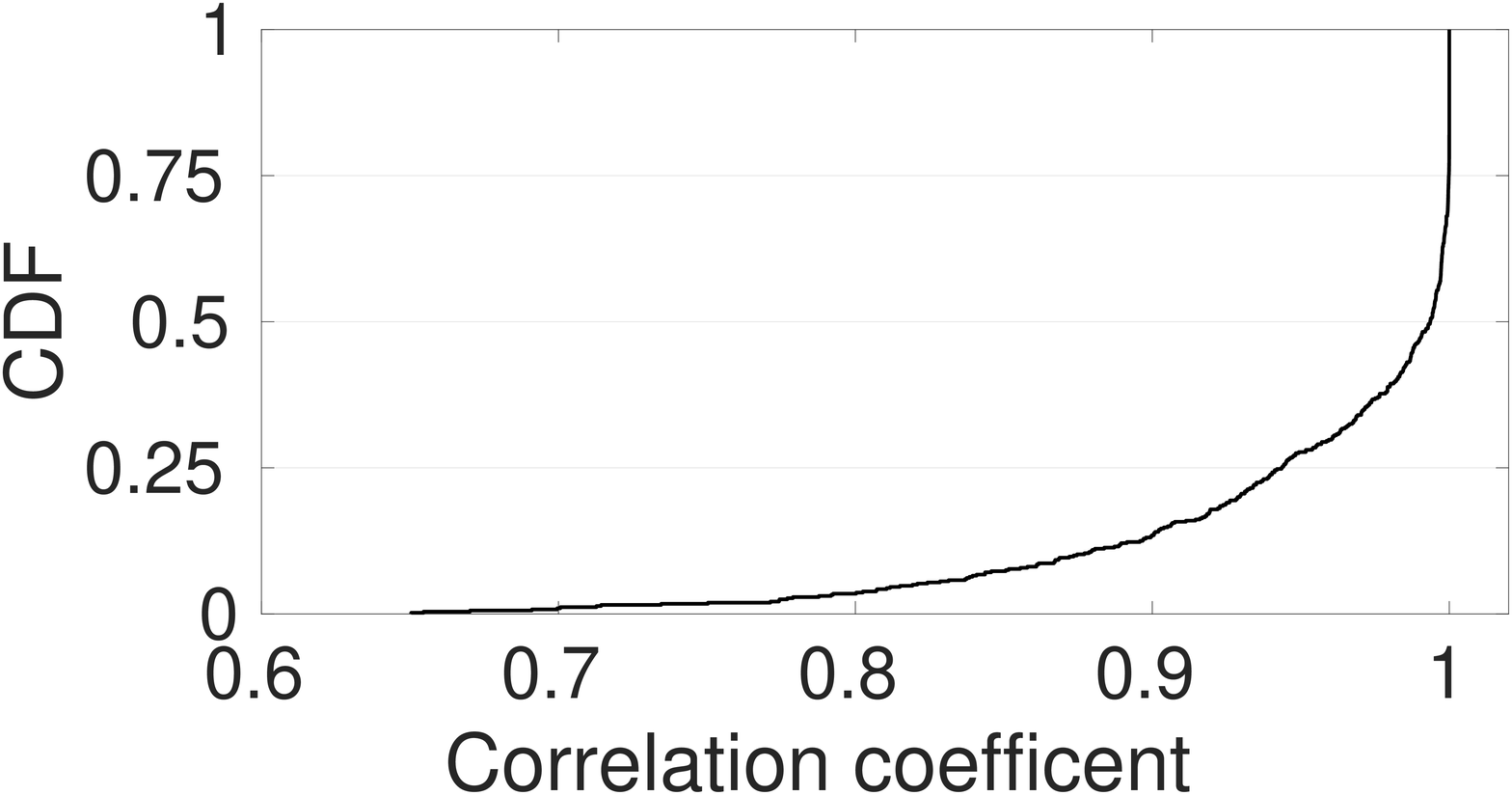}
    \caption{Spatial interference correlation} \label{fig:spatialInterferenceCorrelation}
    \vspace*{-0.1in}
  \end{wrapfigure}

\subHeading{Agile model instantiation for new links.}
Due to vehicle mobility and starting of vehicles, \emph{new links} may form when vehicles come within one another's communication ranges. The need for reliable inter-vehicle communication makes it desirable for the gPRK model parameters of the newly-formed links to quickly converge to their safe-state where application-required link reliabilities are ensured.
    To this end, it is desirable to initialize the gPRK model parameters of newly formed links close to where their safe-state may be, and we propose to leverage \emph{spatial interference correlation} to accomplish this. 
    More specifically, in large-scale wireless networks such as V2V networks, close-by links whose senders and receivers are close to one another respectively tend to experience similar interference power and similar set of close-by, strong interferers, especially if the radii of their receiver-side exclusion regions (ERs) are similar. 
    For the network setting of Section~\ref{sec:eval}, for instance, Figure~\ref{fig:spatialInterferenceCorrelation} shows the empirical
    cumulative distribution function (CDF) of the correlation coefficient between the receiver-side interference power of any two links for which the inter-sender distance, the inter-receiver distance, and the difference in the radii of receiver-side ERs are no more than 30 meters. We see that the correlation coefficient tends to be large.
    This spatial interference correlation enables us to develop mechanisms for accurate gPRK model initialization as below.

When a new link from $S_i$ to $R_i$, denoted by $\langle S_i, R_i \rangle$, is formed at time $t$, $R_i$ first checks whether there exists another sender vehicle $S_j (j \ne i)$ for which the gPRK model parameter $K_{S_j,R_i,T_{S_j,R_i}}(t)$ has converged to a safe state for link $\langle S_j, R_i \rangle$ (i.e., the communication reliability from $S_j$ to $R_i$ has met the requirement $T_{S_j,R_i}$). For convenience, we call the link $\langle S_j, R_i \rangle$ a \emph{self-reference link} for $\langle S_i, R_i \rangle$. Let $\mathbb{S} = \{S_j: \langle S_j, R_i \rangle \ \text{is a self-reference link for} \ \langle S_i, R_i \rangle \}$, and let $S^*$ be the vehicle that is closest to $S_i$ out of all the vehicles in $\mathbb{S}$.
    $R_i$ uses $\langle S^*, R_i \rangle$ to initialize the gPRK model of $\langle S_i, R_i \rangle$ as follows: \emph{$R_i$ first sets the gPRK model parameter of $\langle S_i, R_i \rangle$ such that the ER of $\langle S_i, R_i \rangle$ is the same as that of $\langle S^*, R_i \rangle$, and, based on the assumption that $R_i$ experiences similar interference power when senders $S^*$ and $S_i$ transmit to $R_i$ with the same ER around $R_i$ (i.e., $D(S^*, R_i) K_{S^*,R_i,T_{S^*,R_i}} = D(S_i, R_i) K_{S_i,R_i,T_{S_i,R_i}}$), $R_i$ then uses the gPRK-model-adaptation rules to adjust the model parameter of $\langle S_i, R_i \rangle$ to accommodate the differences between $\langle S_i, R_i \rangle$  and $\langle S^*, R_i \rangle$}.
More specifically, $R_i$ first sets $K_{S_i,R_i,T_{S_i,R_i}}(t) = \frac{D(S^*, R_i,t) K_{S^*,R_i,T_{S^*,R_i}}(t)}{D(S_i, R_i, t)}$, where $D(V_j, V_i, t)$ denotes the geometric distance between two vehicles $V_j$ and $V_i$ at time $t$;
    \sloppy{then $R_i$ computes $\Delta I_{R_i}(t) = P(S_i, R_i, t) - P(S^*, R_i, t) + P(S_i, R_i, t)(\frac{1}{f^{-1}(T_{S_i,R_i})} - \frac{1}{f^{-1}(T_{S^*,R_i})})$, where $P(V_j, V_i, t)$ denotes the signal power from vehicle $V_j$ to $V_i$ at time $t$, the term $P(S_i, R_i, t) - P(S^*, R_i, t)$ accounts for the difference in tolerable interference due to different signal power from $S^*$ and $S_i$, and the term $P(S_i, R_i, t)(\frac{1}{f^{-1}(T_{S_i,R_i})} - \frac{1}{f^{-1}(T_{S^*,R_i})})$ accounts for the change in tolerable interference when the communication reliability requirement by $\langle S_i, R_i \rangle$ changes from $T_{S^*,R_i}$ to $T_{S_i,R_i}$; } 
    finally $R_i$ applies the gPRK-model-adaptation rules 
    (as discussed in Section~\ref{subsec:gPRK}) to adjust the value of $K_{S_i,R_i,T_{S_i,R_i}}(t)$, and the final value of $K_{S_i,R_i,T_{S_i,R_i}}(t)$ is set as the initial gPRK model parameter for the newly formed link $\langle S_i, R_i \rangle$.

If there exists no self-reference link for $\langle S_i, R_i \rangle$ when it newly forms (e.g., when vehicle $R_i$ just got started), $R_i$ tries to identify a \emph{neighbor-reference link} $\langle S_j, R_j \rangle (j \ne i)$ such that the gPRK model parameter $K_{S_j,R_j,T_{S_j,R_j}}(t)$ has converged to a safe state, and $D(S_j, S_i, t)$ as well as $D(R_j, R_i, t)$ are less than a threshold $D_0$, where $D_0$ is chosen such that links $\langle S_j, R_j \rangle$ and $\langle S_i, R_i \rangle$ experience similar interference power when the radii of their ERs are the same (i.e., $D(S_j, R_j) K_{S_j,R_j,T_{S_j,R_j}} = D(S_i, R_i) K_{S_i,R_i,T_{S_i,R_i}}$). Let $\mathbb{L} = \{\langle S_j, R_j \rangle: \langle S_j, R_j \rangle \ \text{is a neighbor-reference link for} \ \langle S_i, R_i \rangle \}$,  define the distance between two links $\langle S_j, R_j \rangle$ and $\langle S_i, R_i \rangle$ at time $t$ as $\max \{D(S_j, S_i, t), D(R_j, R_i, t)\}$, and let $\langle S^*, R^* \rangle$ be the link closest to $\langle S_i, R_i \rangle$ among all the links in
\ifthenelse{\boolean{short}}
{$\mathbb{L}$.}
{$\mathbb{L}$ (i.e., $\langle S^*, R^* \rangle = \argmin_{\langle S_j, R_j \rangle \in \mathbb{L}} \max \{D(S_j, S_i, t), D(R_j, R_i, t)\}$) .}
    $R_i$ then uses $\langle S^*, R^* \rangle$ to initialize the gPRK model for
    \ifthenelse{\boolean{short}}
    {$\langle S_i, R_i \rangle$ as in the case of estimation via self-reference links as discussed above. }
    {$\langle S_i, R_i \rangle$. More specifically, assuming that $R^*$ and $R_i$ experience similar interference power when $S^*$ and $S_i$ transmit respectively with similar communication reliability requirements, $R_i$ sets $K_{S_i,R_i,T_{S_i,R_i}}(t) = K_{S^*,R^*,T_{S^*,R^*}}(t)$ and computes $\Delta I_{R_i}(t) = P(S_i, R_i, t) - P(S^*, R^*, t) + P(S_i, R_i, t)(\frac{1}{f^{-1}(T_{S_i,R_i})} - \frac{1}{f^{-1}(T_{S^*,R^*})})$, then $R_i$ applies the gPRK model adaptation rule Rule-ER0, Rule-ER1, or Rule-ER2 to adjust the value of $K_{S_i,R_i,T_{S_i,R_i}}(t)$, and the final value of $K_{S_i,R_i,T_{S_i,R_i}}(t)$ is set as the initial gPRK model parameter for the newly formed link $\langle S_i, R_i \rangle$.}

    Leveraging the spatial correlation between $\langle S_i, R_i \rangle$ and its self-reference and neighbor-reference links, the above gPRK model initialization mechanism enables good approximation of the safe-state gPRK model parameter of $\langle S_i, R_i \rangle$ in normal and heavy vehicle traffic settings where there are usually enough number of surrounding vehicles/links around $\langle S_i, R_i \rangle$.
    In the case of very light vehicle traffic settings (e.g., at 3 a.m$.$), there may exist no self-reference link nor neighbor-reference link for a newly formed link $\langle S_i, R_i \rangle$. 
In this case, vehicles are sparsely distributed, cumulative interference from far-away vehicles tends to be small, and the exclusion region (ER) tends to be smaller than in the case of normal and heavy vehicle traffic settings. Accordingly, $R_i$ can approximate its safe-state gPRK model parameter by only considering pairwise interference among close-by vehicles. More precisely, $R_i$ sets the initial value of the gPRK model parameter such that the initial ER around itself includes every vehicle whose transmission alone, concurrent with the transmission from $S_i$ to $R_i$, can make the communication reliability drop below $T_{S_i,R_i}$.

\subHeading{Agile model instantiation for transient links.}
For an \emph{established link} $\langle S_i, R_i \rangle$ where $S_i$ and $R_i$ are on different roads or in opposite driving directions of the same road, the link may be \emph{transient} since the relative position between $S_i$ and $R_i$ and thus the link properties between them may change significantly during an update interval of the gPRK model parameter (e.g., every 2.5 seconds). In this case, the gPRK-model-adaptation rules of Section~\ref{subsec:gPRK} won't be agile enough to track the gPRK model parameter of $\langle S_i, R_i \rangle$.
    Thus we propose to treat the transient link between $S_i$ and $R_i$ as a ``new" link from one time slot to the next and use the gPRK model initialization approach presented above to instantiate the gPRK model parameter of $\langle S_i, R_i \rangle$.
    In normal and heavy vehicle traffic settings, vehicles of the same traffic flow (i.e., vehicle traffic along the same direction of a road segment) tend to form \emph{clusters} depending on their speed, with the vehicles in the same cluster having approximately equal speed and relatively stable spatial distribution, and this clustering behavior applies to both free-flow and congested traffic and for both highways and urban roads
    \ifthenelse{\boolean{short}}
    {\cite{Martin:trafficFlowDynamicsBook}.}
    {\cite{Gavrilov:free-flow-traffic-structure,Helbing:congested-traffic-patterns,Panichpapiboon:density-estimation,Martin:trafficFlowDynamicsBook,Ho:Urban-traffic-Poisson,Bai:highway-Poisson-distribution}.}
    With spatiotemporal constraints on vehicle movement along a traffic flow, vehicle cluster membership tends to last at timescales from seconds to minutes or even longer
    \ifthenelse{\boolean{short}}
    {\cite{Martin:trafficFlowDynamicsBook}.}
    {\cite{Gavrilov:free-flow-traffic-structure,Helbing:congested-traffic-patterns,Panichpapiboon:density-estimation,Martin:trafficFlowDynamicsBook,Ho:Urban-traffic-Poisson,Bai:highway-Poisson-distribution}.}
    The relative stability of intra-cluster vehicle spatial distribution and cluster membership make the gPRK-model-adaptation rules applicable to the links between vehicles of the same cluster, and these stable links can serve as the self-reference and neighbor-reference links for other transient links, thus enabling online, adaptive instantiation of the gPRK model parameters of transient links.
In the case of very light vehicle traffic where there may exist no self-reference nor neighbor-reference link for a transient link $\langle S_i, R_i \rangle$, the gPRK model parameter of $\langle S_i, R_i \rangle$ may be instantiated by considering pairwise interference as discussed earlier.

Another type of \emph{transient link} $\langle S_i, R_i \rangle$ exists when $R_i$ repeatedly moves in and out of the communication range of $S_i$. In this case, if the interval between $R_i$ moving out of and then back into the communication range of $S_i$ is small (e.g., less than 2 seconds), then $\langle S_i, R_i \rangle$ can retain its last gPRK model parameter considering the \emph{temporal correlation of interference} at the receiver
\ifthenelse{\boolean{short}}
{$R_i$ (as we elaborate in more detail in \cite{CPS-TR});}
{$R_i$;}
if the interval is large,  $\langle S_i, R_i \rangle$ can be treated as a new link, and its gPRK model parameter can be initialized using the gPRK model initialization method discussed earlier.
\ifthenelse{\boolean{short}}
{}
{
For the network setting of Section~\ref{sec:eval} and a time lag of 2 seconds, Figure~\ref{fig:temporalInterferenceCorrelation} shows the empirical CDF
  \begin{wrapfigure}{r}{.5\linewidth}
    \vspace*{-0.1in}
    \centering
    \includegraphics[width=\linewidth]{temporal-interference-correlation-lessthan-2sec}
    \caption{Temporal interference correlation} \label{fig:temporalInterferenceCorrelation}
    \vspace*{-0.1in}
  \end{wrapfigure}
   of the auto-correlation coefficient of receiver-side interference power. We see that the correlation coefficient tends to be large, thus validating the aforementioned assumption of temporal interference correlation.
}

\subHeading{Effective use of gPRK model.}
In order for vehicles to use the gPRK model to detect their mutual interference relations in a distributed manner, close-by, potentially interfering vehicles need to be aware of one another's locations. A vehicle can update and share its location with close-by vehicles by broadcasting its location periodically. In the presence of high vehicle mobility, however, the relative positions of two vehicles may change in an non-negligible manner during the broadcast intervals. For instance, even if the location information is updated every half a second, the distance between two vehicles driving at a speed of 80km/h (i.e., 50mph) along the opposite directions of a road may change 22.22 meters during the update interval.
    In order for vehicles to have accurate information about one another's locations during update intervals and with limited location update frequencies, we propose to have vehicles estimate one another's locations during update intervals. For accurate estimation of vehicle locations, it is important to have a good model for vehicle location dynamics.

Fortunately, vehicle dynamics have been studied extensively in traffic flow theory, and the intelligent-driver-model (IDM) as well as its extensions have been shown to be able to accurately model real-world, microscopic vehicle dynamics \cite{Martin:trafficFlowDynamicsBook}.
\ifthenelse{\boolean{short}}
{Using the IDM model and by treating vehicle location as a part of the ``state" of a vehicle, we can derive the dynamic model of the vehicle. (Details of the derivation can be found in \cite{CPS-TR}.)}
{A key part of the models is the model on vehicle acceleration behavior according to the speed and locations of a vehicle and its surrounding vehicles. In this study, we use an enhanced version of IDM which captures precisely the behavior of adaptive cruise control (ACC). In the model, the vehicle acceleration function $a_{ACC}$ is defined by (\ref{eqn:a-ACC}) below \cite{Martin:trafficFlowDynamicsBook}:
{\small
\begin{equation} \label{eqn:a-ACC}
a_{ACC}(s, v, v_l, \dot{v_l}) = \begin{cases}
	a_{IIDM},                                                                  & \textrm{if} \ a_{IIDM} \ge a_{CAH}  \\
	(1-c)a_{IIDM} + c[a_{CAH} +   \                             &  \textrm{otherwise} \\
   b \textrm{tanh}(\frac{a_{IIDM} - a_{CAH}}{b}) ],  &   \\
	\end{cases}
\end{equation}
}
where $c \in [0, 1]$ and is usually set as $0.99$,
{\small
\begin{equation} \label{eqn:a-CAH}
a_{CAH}(s, v, v_l, \dot{v_l}) = \begin{cases}
	\frac{v^2 \tilde{a_l}}{v_l^2 - 2 s \tilde{a_l}},        & \textrm{if} \ v_l(v-v_l) \le -2s\tilde{a_l}  \\
	\tilde{a_l} - \frac{(v-v_l)^2 I_{v-v_l \ge 0}}{2s},  \  &  \textrm{otherwise}
	\end{cases}
\end{equation}

\begin{equation} \label{eqn:a-IIDM}
a_{IIDM} = \begin{cases}
	a(1-z^2),                               &                 \textrm{if} \ v \le v_0, z=\frac{s^*(v, v-v_0)}{s} \ge 1  \\
	a_{free}(1 - z^{(2a)/a_{free}}),  \  &  \textrm{if}  \ v \le v_0, z < 1  \\
	a_{free} + a(1-z^2),                           &  \textrm{if} \ v > v_0, z \ge 1  \\
                    a_{free},                                               &  \textrm{if} \ v > v_0, z < 1
	\end{cases}
\end{equation}

\begin{equation} \label{eqn:a-free}
a_{free}(v) = \begin{cases}
	a[1- (\frac{v}{v_0})^\delta],                               & \textrm{if} \ v \le v_0  \\
	-b[1- (\frac{v}{v_0})^{a\delta/b}],  \                &  \textrm{otherwise}
	\end{cases}
\end{equation}

\begin{equation} \label{eqn:s-star}
s^*(v, v-v_0) = s_0 + \max (0, vT + \frac{v(v - v_l)}{2\sqrt{ab}}).
\end{equation}
}
In the above equations, $v$ and $v_l$ represent the speed of the modeled vehicle and its lead vehicle (i.e., the vehicle immediately preceding the vehicle considered) respectively, $s$ is the rear-bumper-to-front-bumper distance from the lead vehicle to the vehicle, $\dot{v_l}$ is the acceleration of the lead vehicle, and the effective acceleration of the lead vehicle used in modeling is $\tilde{a_l} = \min(\dot{v_l}, a)$. In the above model, the parameters $v_0, T, s_0, \delta, a$, and $b$ represent the desired speed, time gap between the vehicle and its lead vehicle, minimum space gap between the vehicle and its lead vehicle, acceleration exponent, maximum acceleration, and comfortable deceleration respectively \cite{Martin:trafficFlowDynamicsBook}.

Using the above model and by treating the speed, location, and bumper-to-bumper distance to its lead vehicle as the ``state'' of a vehicle, we can derive the dynamic model of the vehicle.} 
Given that the model is nonlinear, we use the Unscented Kalman Filter (UKF) \cite{Wan:UKF-for-nonlinear} to estimate vehicle locations. By treating the model parameters as a part of the system state and introducing random walks to the parameter evolution \cite{Wan:UKF-for-nonlinear}, the microscopic model can also be adapted according to the individual driving behavior of vehicles in different real-world settings. 
Besides vehicle location estimation, the above approach to vehicle location estimation can be applied to a vehicle itself to filter out its own location measurement errors for improved localization accuracy.

The IDM model focuses on the longitudinal movement of a vehicle along a specific lane, and it does not directly model behavior such as lane change and turn. Since it is more difficult to model those behavior accurately \cite{Martin:trafficFlowDynamicsBook}, we propose, for effectiveness of real-world deployment, not to explicitly model those behavior and resort to event-based quick diffusion of vehicle state to address the impact of lane change and turn; that is, a vehicle immediately shares its new location right after it changes lane or turns. 
Together, these mechanisms enable vehicles to be aware of one another's locations, thus enabling the effective use of the gPRK model in V2V networks.

\ifthenelse{\boolean{ieee10pager}}
{}
{
\section{Supporting Predictable Broadcast}  \label{sec:addressBroadcast}

\subHeadingS{Sender ER for reliable broadcast.}
A fundamental communication primitive in V2V networks is single-hop broadcast via which a vehicle shares its state (e.g., location and speed) with close-by vehicles within a certain distance \cite{Zhang:V2X-survey}. Reliable broadcast is a well-known challenge because, even though acknowledgments from receivers are required for many reliability-enhancement mechanisms such as ACK-/negative-ACK-based retransmission of lost packets and RTS-CTS-based collision avoidance in medium access control, it is difficult for a sender to reliably and efficiently get an acknowledgment from every receiver, especially when the number of receivers is large in V2V networks (e.g., up to hundreds).

To address the challenge, we observe that, to ensure a minimum broadcast reliability $T_S$ for a sender $S$, we need to make sure that the communication reliability along the link from  $S$ to every one of its receiver $R_i \in \mathbf{R}$ is at least $T_S$. This fact enables us to define, for a broadcast sender $S$, a \emph{receiver exclusion region (ER)} $\mathbb{E}_{S,R_i,T_S}$ for every receiver $R_i \in \mathbf{R}$ based on the gPRK model. Accordingly, we define the \emph{sender ER} for $S$, denoted by $\mathbb{E}_{S,T_S}$,  as the union of
\begin{wrapfigure}{r}{.4\linewidth}
    \vspace*{-0.1in}
    \centering
    \includegraphics[width=0.9\linewidth]{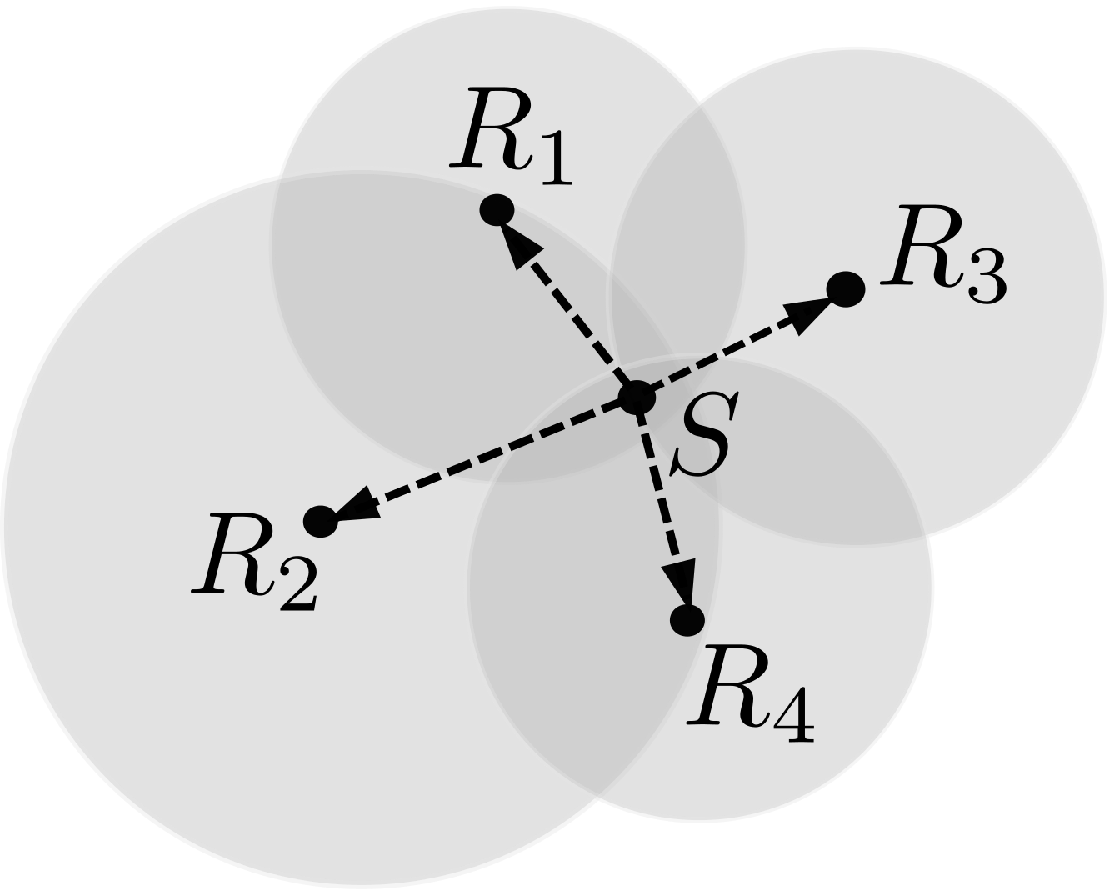}
    \caption{Sender ER} \label{fig:senderER}
    \vspace*{-0.1in}
\end{wrapfigure}
its corresponding receiver ERs; that is, $\mathbb{E}_{S,T_S} = \cup_{R_i \in \mathbf{R}} \mathbb{E}_{S,R_i,T_S}$. For instance, Figure~\ref{fig:senderER} shows an example of the sender ER when the sender $S$ has four receivers $R_1, R_2, R_3$ and $R_4$.
    Based on the definition of the sender ER, the broadcast reliability of $T_S$ is ensured as long as no node in $\mathbb{E}_{S,T_S}$ transmits concurrently with sender $S$.

\ifthenelse{\boolean{short}}
{}
{
As we have discussed in Section~\ref{subsec:gPRK}, the gPRK model uses inter-vehicle distance 
to identify interference relations between vehicles. Thus the ER around a receiver $R_i$ in the gPRK model assumes a disk shape instead of a potentially irregular geometric shape in the PRK model, as shown in Figure~\ref{fig:gPRK-vs-PRK-receiverER}.
    \begin{figure}[!htbp]
    \centering
    \begin{subfigure}[b]{0.2\textwidth}
        \centering
        \includegraphics[width=.99\textwidth]{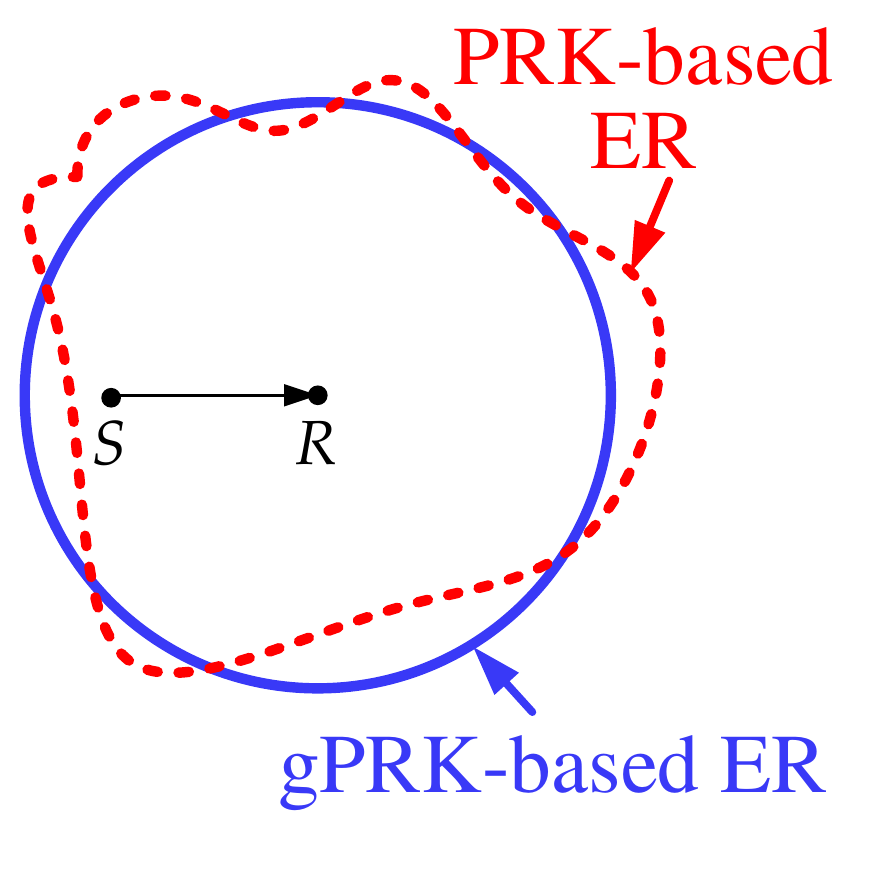}
        \caption{gPRK- vs$.$ PRK-based receiver ER}  \label{fig:gPRK-vs-PRK-receiverER}
    \end{subfigure}%
    \ \ \
    \begin{subfigure}[b]{0.25\textwidth}
        \centering
        \includegraphics[width=.99\textwidth]{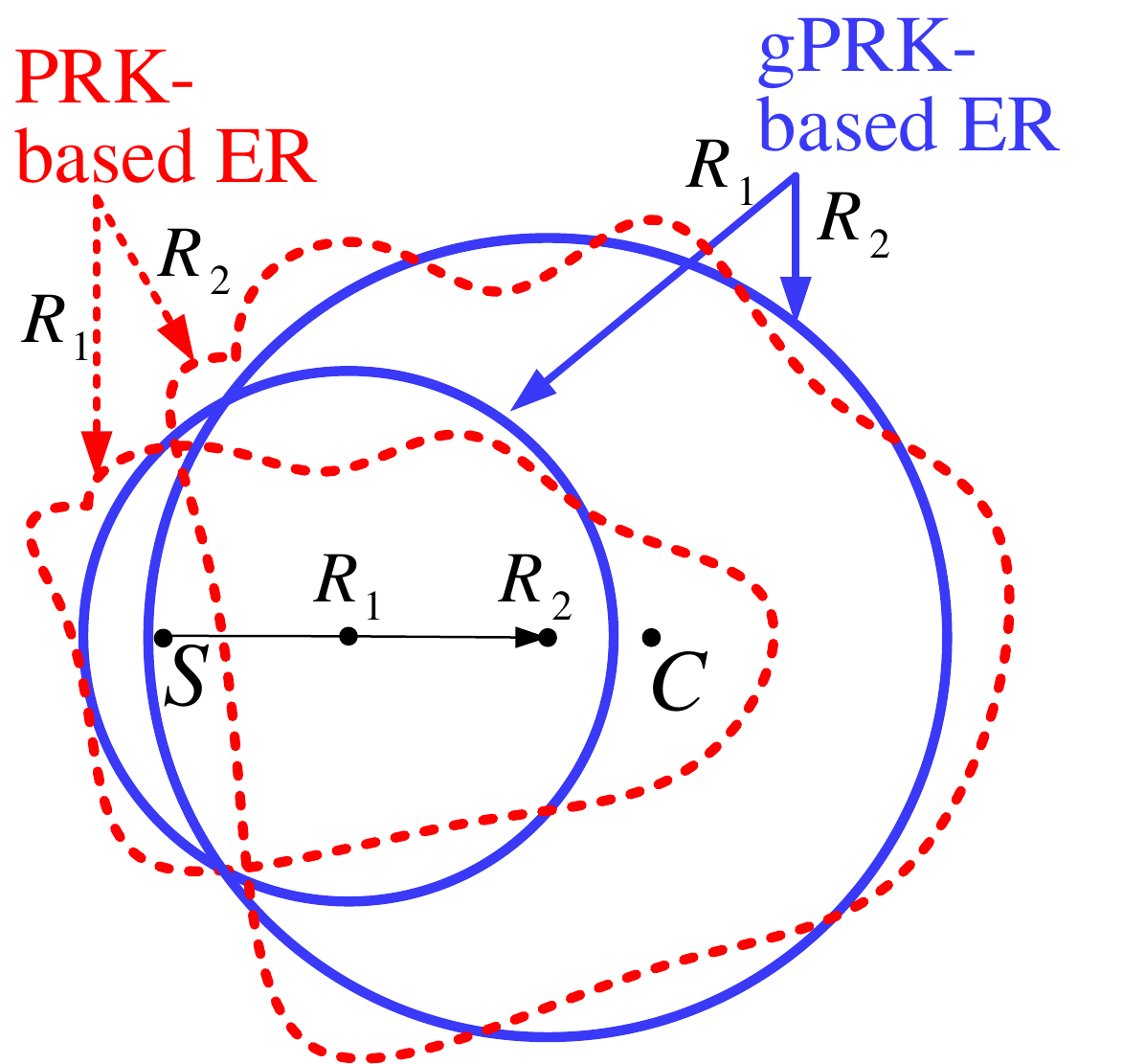}
        \caption{gPRK- vs$.$ PRK-based sender ER}  \label{fig:gPRK-vs-PRK-senderER}
    \end{subfigure}
    \caption{gPRK- vs. PRK-based exclusion region (ER)}
\end{figure}
Accordingly, the set of vehicles inside the ER of $R_i$ may be different in the gPRK and PRK models. For links with high communication reliability requirements in large scale networks such as V2V networks, the receiver ERs tend to be large (e.g., with a geometric radius more than twice the sender-receiver distance in many scenarios \cite{PRK}), and thus the number of differing vehicles in the gPRK- and PRK-based ERs tends to be relatively small as compared to the total number of vehicles in the ERs.
    For inter-vehicle broadcast, since the sender ER of a vehicle $S$ is the union of the ERs around all of its receivers, the size of the sender ER is even larger than the size of individual receiver ERs, and a vehicle in the interior of the sender ER that is in the PRK-based but not in the gPRK-based ER (or in the gPRK-based but not in the PRK-based ER) of a receiver $R_i$ may well be in the gPRK-based ER (or the PRK-based ER) of another receiver $R_j (j \ne i)$. Therefore, the differences between gPRK- and PRK-based sender ERs tend to be even less significant. As shown in Figure~\ref{fig:gPRK-vs-PRK-senderER}, for instance, vehicle $C$ is in the PRK-based ER but not in the gPRK-based ER of receiver $R_1$, but $C$ is in the gPRK-based ER of $R_2$, thus $C$ is in both the PRK- and gPRK-based sender ER of $S$.
\ifthenelse{\boolean{temp}}
{}
{???These observations are corroborated in Section~\ref{sec:eval} where we experimentally show that gPRK-based scheduling enables statistically the same concurrency as PRK-based scheduling while ensuring the application-required communication reliability. }
} 

\subHeading{Broadcast receiver ER adaptation.}
For reliable inter-vehicle broadcast, a vehicle $C$ in the sender ER of another vehicle $S$ shall not transmit concurrently with $S$. Given that the sender ER of a vehicle $S$ is the union of the receiver ERs of $S$'s receivers, a vehicle $C$ may lie in the receiver ER of multiple receivers; that is, the receiver ERs of two receivers $R_i$ and $R_j$ ($i \ne j$) may overlap and share some common vehicles. 
    The overlap of receiver ERs and the fact that the sender ER is the union of all receiver ERs make the adaptation of gPRK model parameters and thus ERs around individual receivers interact with one another. To ensure predictable broadcast reliability in the presence of network dynamics such as vehicle mobility, we need to address the \emph{interaction} between the adaptation of ERs around individual receivers of the same sender. 
In particular, after a receiver $R_i$ computes $\Delta I_{R_i}(t)$ at time $t$ and if $\Delta I_{R_i}(t) \ne 0$, $R_i$ needs to consider whether a vehicle $C$ lies in the receiver ER of another receiver $R_j (j\ne i)$ when deciding to add or remove $C$ from the receiver ER of $R_i$ itself.

When $\Delta I_{R_i}(t) < 0$ (i.e., $R_i$ needs to expand its receiver ER), the receiver ER expansion rule Rule-ER1 needs to be amended with the following rule:
\begin{itemize}
\item[] \ruleHeading{BC1}
If a vehicle $C$ is not in the receiver ER of $R_i$ but is in the receiver ER of another receiver $R_j (j \ne i)$ at time $t$ (i.e., $C \in \mathbb{E}_{S,R_j,T_S}(t) \setminus \mathbb{E}_{S,R_i,T_S}(t))$, $R_i$ treats $I(C,R_i,t)$ as zero.
\end{itemize}
The rationale for Rule-BC1 is that, if $C$ is already in the receiver ER of another receiver $R_j$, $C$ is already in the sender ER of $S$ and does not transmit concurrently with the broadcast transmission by $S$, and thus $C$ does not introduce any interference to the receiver $R_i$ and its effective interference power to $R_i$ (i.e., $I(C,R_i,t)$ is zero.

When $\Delta I_{R_i}(t) > 0$ (i.e., $R_i$ needs to shrink its receiver ER), the receiver ER shrinking rule Rule-ER2 needs to be amended with Rule-BC1 for the same rationale as discussed above. However, if the receiver ER of $R_i$ is completely covered by other receivers' ERs at time $t$ (i.e., $\mathbb{E}_{S,T_S}(t) = \mathbb{E}_{S,T_S}(t) \setminus \mathbb{E}_{S,R_i,T_S}(t)$) or if applying Rule-ER2 and Rule-BC1 at time $t$ would make the receiver ER of $R_i$ completely covered by other receivers' ERs, Rule-ER2 and Rule-BC1 cannot be directly applied since applying these rules would lead to an empty receiver ER for $R_i$, which could make the communication reliability to $R_i$ unpredictable in the presence of network dynamics as we will discuss shortly.
    In this case, we regard $R_i$ as an \emph{unconstrained receiver} of $S$ at time $t$ since the sender ER of $S$ will only depend on the receiver ERs of those receivers other than $R_i$. Accordingly, we regard a receiver $R_j$ as a \emph{constrained receiver} of $S$ at time $t$ if $R_j$ is not an unconstrained receiver.

For an unconstrained receiver $R_i$ at time $t$, its receiver ER does not impact the sender ER $\mathbb{E}_{S,T_S}(t)$ at time $t$, thus we could arbitrarily set its ER if we did not consider network dynamics such as vehicle mobility. Due to vehicle mobility, however, the set of vehicles whose receiver ERs jointly cover that of $R_i$ at time $t$ may move such that their receiver ERs do not cover that of $R_i$ at time $t+1$. 
To address the impact of network dynamics, we propose the following rule of adapting the receiver ER of an unconstrained receiver $R_i$ so that \emph{the communication reliability from $S$ to $R_i$ is still ensured at time $t+1$ even if network dynamics (e.g., vehicle mobility) is such that $R_i$'s receiver ER is not covered by others' receiver ERs at time $t+1$}. (Note that, for a constrained receiver $R_j$ with $\Delta I_{R_j}(t) > 0$, Rule-ER2 and Rule-BC1 apply.)
    \begin{itemize}
      \item[] \ruleHeading{BC2}

      (A) If $\Delta I_{R_i}(t) > 0$, $R_i$ is an unconstrained receiver of $S$, and the receiver ER of $R_i$ is completely covered by other receivers' ERs at time $t$ (i.e., $\mathbb{E}_{S,T_S}(t) = \mathbb{E}_{S,T_S}(t) \setminus \mathbb{E}_{S,R_i,T_S}(t)$), $R_i$ expands its receiver ER to the largest possible that is still completely covered by other receivers' ERs (i.e., sets $K_{S,R_i,T_S}(t)$ to the largest value that still ensures $\mathbb{E}_{S,T_S}(t) = \mathbb{E}_{S,T_S}(t) \setminus \mathbb{E}_{S,R_i,T_S}(t)$), and then $R_i$ applies Rule-ER2 (but not Rule-BC1) to shrink its receiver ER.

      (B) If $\Delta I_{R_i}(t) > 0$, $R_i$ is an unconstrained receiver of $S$, and the receiver ER of $R_i$ is not completely covered by other receivers' ERs at time $t$ (i.e., $\mathbb{E}_{S,T_S}(t) \ne \mathbb{E}_{S,T_S}(t) \setminus \mathbb{E}_{S,R_i,T_S}(t)$), $R_i$ first lets $\mathbb{E}_0 = \mathbb{E}_{S,R_i,T_{S,R}}(t)$, then keeps removing nodes out of $\mathbb{E}_{S,R_i,T_{S,R}}(t)$, in the non-increasing order of their distance to $R$, until the condition $\mathbb{E}_{S,T_S}(t) = \mathbb{E}_{S,T_S}(t) \setminus \mathbb{E}_{S,R_i,T_S}(t)$ holds for the first time. Then $R_i$ sets $\Delta I_{R_i}(t)$ as $\Delta I_{R_i}(t) - \sum_{C \in \mathbb{E}_0 \setminus \mathbb{E}_{S,R_i,T_{S,R}}(t)} I(C, R, t)$, where $I(C, R, t)$ is computed in conformance with Rule-BC1. Then $R_i$ applies Rule-ER2 (but not Rule-BC1) to shrink its receiver ER.
    \end{itemize}

In Rule-BC2(A), the reason why $R_i$ first expands its receiver ER to the largest possible that is still completely covered by other receivers' ERs is to make sure that, before applying Rule-ER2,  the value of $\mathbb{E}_{S,R,T_{S,R}}(t)$ corresponds to the network setting 
from which the value of $\Delta I_{R_i}(t)$ is derived while pretending 
that the receiver ER of $R_i$ was not covered by others' receiver ERs.

 With Rule-BC2, the communication reliability from $S$ to $R_i$ is ensured even if the receiver ER of $R_i$ is not covered by others' receiver ERs at time $t+1$. This property is important for V2V networks with high vehicle mobility.
A special case is when a vehicle $R_j (j \ne i)$ at the boundary of the broadcast communication range of $S$ moves outside the communication range of $S$ while $R_i$ is the the next vehicle closest to the boundary of $S$'s communication range, as shown in
Figure~\ref{fig:Rule-BC2}. In this case, 
$R_i$'s receiver ER is covered by that of $R_j$, and a significant portion of $S$'s sender
  \begin{wrapfigure}{r}{.4\linewidth}
    \vspace*{-0.1in}
    \centering
    \includegraphics[width=0.99\linewidth]{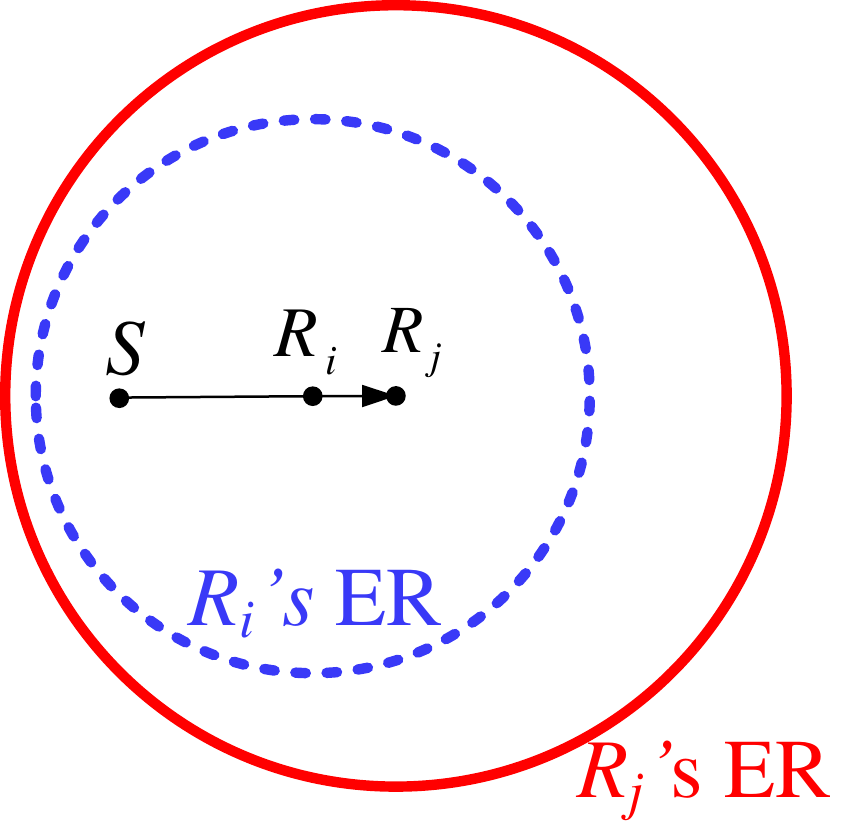}
    \caption{Benefit of Rule-BC2} \label{fig:Rule-BC2}
    \vspace*{-0.2in}
  \end{wrapfigure}
ER is also covered by $R_j$'s receiver ER.\footnote{For clarity of Figure~\ref{fig:Rule-BC2}, the figure does not show other receivers of $S$ nor their ERs.} With Rule-BC2, $R_i$ sets
its receiver ER by pretending that it was not covered by that of $R_j$, thus the communication reliability from $S$ to $R_i$ 
is ensured when $R_j$ moves outside the communication range of $S$; without Rule-BC2, the receiver ER around $R_i$ would become empty, which would make the communication reliability from $S$ to $R_i$  unpredictable when $R_j$ moves outside the communication range of $S$.

\ifthenelse{\boolean{short}}
{}
{
\subHeading{Lightweight signaling of sender ER.}
Two vehicles $S_i$ and $S_j$ ($i \ne j$) interfere with each other and thus cannot transmit concurrently if $S_i \in \mathbb{E}_{S_j, T_{S_j}}$ and/or $S_j \in \mathbb{E}_{S_i, T_{S_i}}$, where $T_{S_i}$ and $T_{S_j}$ are the broadcast reliability requirements by $S_i$ and $S_j$ respectively. In order for vehicles to know the mutual interference relations among themselves, each vehicle $S$ needs to share its sender ER $\mathbb{E}_{S, T_{S}}$ with potentially interfering vehicles. Since the sender ER is the union of all receiver ERs and the number of broadcast receivers may be large (e.g., up to hundreds), the overhead for a sender to signal its sender ER with potentially interfering vehicles will be high if we represent the sender ER by listing all the receiver ERs individually. High signaling overhead not only reduces effective network capacity, it also makes it difficult for vehicles to share their sender ERs in a timely manner and to be accurately aware of their mutual interference relations. 
    To address the challenge, we observe that, for a given sender $S$, its receiver ERs tend to overlap with one another, especially in heavy vehicle traffic settings. To minimize signaling overhead, $S$ needs to minimize the number of receiver ERs it signals. By treating each receiver ER as the set of vehicles inside the ER, the minimum signaling overhead problem can be formulated as a \emph{minimum-set-cover (MSC)} problem where the sender $S$ needs to select a minimum number of receiver ERs such that the selected receiver ERs cover all the vehicles that are in the sender ER. More formally, given a sender $S$ and its receivers $\mathbf{R}(t)$ at time $t$, the problem can be formulated as follows: 
    \begin{equation}
    \begin{array}{ll}
    \underline{\text{Problem}}          & \mathbf{P}_{\text{MSC}} \\
    \text{minimize}                            &  |\mathbf{R'}(t)|  \\
    \text{subject to}                           & \mathbf{R'}(t) \subseteq \mathbf{R}(t) \\
                                                         & \cup_{R_i \in \mathbf{R'}(t)}  \mathbb{E}_{S,R_i,T_S}(t) = \cup_{R_i \in \mathbf{R}(t)}  \mathbb{E}_{S,R_i,T_S}(t)  \\
    \end{array}
    \end{equation}
It is well-known that the MSC problem is NP-hard \cite{set-cover}, thus we use the following simple, greedy algorithm to solve problem $\mathbf{P}_{\text{MSC}}$:
    \begin{itemize}
    \item Denote the optimal solution to be $\mathbf{R^*}(t)$, and initialize $\mathbf{R^*}(t)$ to be $\emptyset$;
    \item Iteratively add receivers to $\mathbf{R^*}(t)$ until $\cup_{R_i \in \mathbf{R^*}(t)}  \mathbb{E}_{S,R_i,T_S}(t) = \cup_{R_i \in \mathbf{R}(t)}  \mathbb{E}_{S,R_i,T_S}(t)$; at each step of the iterative process, choose the receiver whose receiver ER contains the largest number of uncovered vehicles.
    \end{itemize}
    The time complexity of the above algorithm is $O(|\mathbf{R}(t)| |\cup_{R_i \in \mathbf{R}(t)}  \mathbb{E}_{S,R_i,T_S}(t)|)$. According to Chvatal \cite{set-cover}, the above greedy algorithm achieves an approximation ratio no larger than $\ln (|\cup_{R_i \in \mathbf{R}(t)}  \mathbb{E}_{S,R_i,T_S}(t)|) + 1$, which tends to be
    \ifthenelse{\boolean{short}}
    {small. Our experimental results in Section~\ref{sec:eval} show that the median and maximum reduction in signaling overhead as a result of the set-cover-based approach is up to 75\% and 97.37\% respectively. }
    {
    small.
    }
Therefore, the above set-cover-based approach enables lightweight signaling of sender ER in a timely manner.

}

  }

\section{Experimental Analysis}  \label{sec:eval}

Considering the lack of large-scale, field-deployed 
V2V network testbeds for evaluating link layer scheduling mechanisms, we implement our CPS scheduling framework in the widely-used ns-3 \cite{ns-3} network simulator, and we experimentally analyze the behavior of CPS by integrating high-fidelity ns-3-based wireless network simulation and SUMO-based vehicle dynamics simulation \cite{SUMO}.

\subsection {Methodology}

\subHeadingS{Multi-dimensional high-fidelity simulation.}
High-fidelity simulation of V2V networks requires high-fidelity simulation of V2V wireless channels and vehicle mobility dynamics. For V2V wireless channels, we implement in ns-3 a channel model based on real-world measurement data that capture large-scale path loss, small-scale fading, and real-world complexities such as anisotropic, asymmetric wireless signal attenuation \cite{v2v-channel-model}.
    For vehicle mobility dynamics, we use the SUMO simulator that simulates 
    vehicle traffic flow dynamics at high-fidelity based on real-world road and traffic conditions of Detroit, Michigan, USA \cite{SUMO}.
    \ifthenelse{\boolean{ieee10pager}}
    {For integrated, high-fidelity simulation of V2V wireless channels and vehicle mobility, we integrate SUMO simulation with ns-3 simulation through the traffic control interface (TraCI) of SUMO \cite{SUMO}.}
    {
    For integrated, high-fidelity simulation of V2V wireless channels and vehicle mobility, we integrate SUMO simulation with ns-3 simulation through the traffic control interface (TraCI) of SUMO, as shown in
    \ifthenelse{\boolean{short}}
    {}
    {\begin{wrapfigure}{r}{.6\linewidth}
    \centering
    \includegraphics[width=0.99\linewidth]{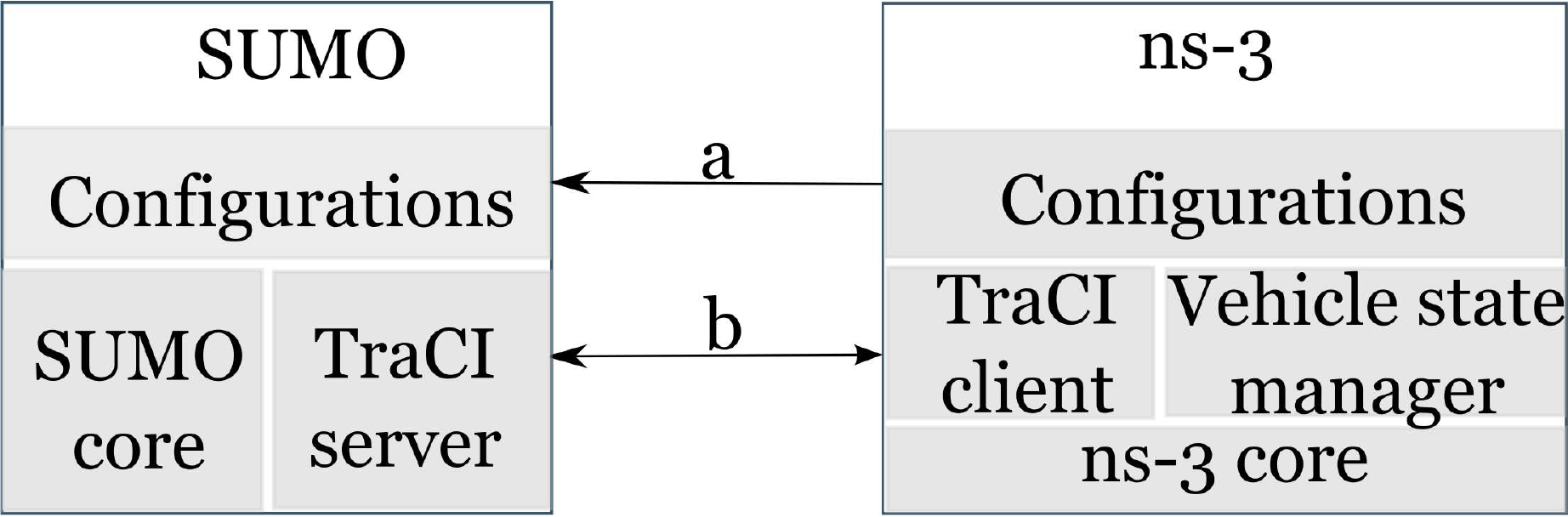}
    \caption{Integration of SUMO with ns-3} \label{fig:sumo_ns3_traci}
    \vspace*{-0.1in}
    \end{wrapfigure}
    }
    Figure~\ref{fig:sumo_ns3_traci}. With the TraCI interface, ns-3 can query any desired
    information (e.g., locations of individual vehicles) from SUMO anytime. When a simulation starts, ns-3 first invokes SUMO with its local configuration files, as shown by link \emph{a} of Figure~\ref{fig:sumo_ns3_traci}; during a
    \ifthenelse{\boolean{short}}
    {\begin{wrapfigure}{r}{.6\linewidth}
    \centering
    \includegraphics[width=0.99\linewidth]{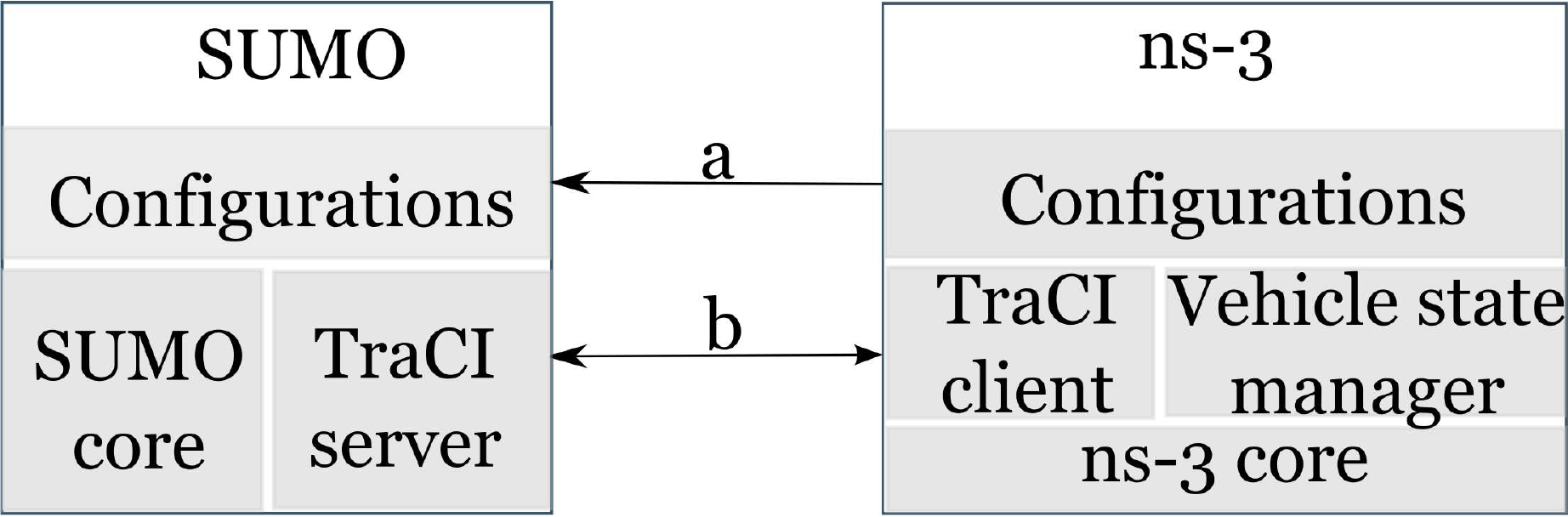}
    \caption{Integration of SUMO with ns-3} \label{fig:sumo_ns3_traci}
    \vspace*{-0.1in}
    \end{wrapfigure}
    }{}
     ns-3 simulation, ns-3 continuously queries vehicle state information (e.g., locations) from SUMO, as shown by link \emph{b} of Figure~\ref{fig:sumo_ns3_traci}. 
     } 

CPS assumes that each vehicle has a location sensor (e.g., GPS and/or SLAM) which reports its real-time locations. To simulate location measurement errors, 
our experimental analysis assumes that the error is a Gaussian variate with zero mean and a standard deviation of four meters, a localization accuracy achievable by today's GPS systems.

\subHeading{Protocols.}
To understand the benefits of CPS in scheduling inter-vehicle communications, we comparatively study the following representative V2V network protocols:
\begin{itemize}
  \item \emph{802.11p}: the MAC protocol of the IEEE 802.11p standard which uses CSMA/CA to coordinate channel access and interference control \cite{Ma:DSRC-reliability-analysis}. This is the MAC protocol used in existing field deployments of DSRC implementations (e.g., those by USDOT).

  \item \emph{DCC}: an ETSI standard that, on top of the 802.11p protocol, uses congestion, power, and rate control to mitigate inter-vehicle interference and improve communication reliability \cite{Xinzhou:VANET-congestion-power-rate-control}.

  \item \emph{AMAC}: the ADHOC MAC protocol \cite{ADHOC-MAC} which is a slot-reservation-based TDMA protocol based on the protocol interference model. In the protocol, vehicles transmit in their reserved slots without carrier sensing. If collisions are detected in a certain time slot of the TDMA frame, vehicles will release the slot and reserve another slot .

  \item \emph{VDDCP}: a TDMA-based MAC protocol \cite{VDDCP} that, based on the protocol interference model, first allocates non-overlapping sets of time slots to different roads and then let vehicles on each road compete for channel access in a slot-reservation-based TDMA manner as in AMAC. 
\end{itemize}

\ifthenelse{\boolean{temp}}
{}
{
???
To understand the capability of CPS in achieving high channel spatial reuse while ensuring application-required broadcast reliability, we also comparatively study the concurrency (i.e., number of concurrent transmissions at a time instant) in a state-of-the-art, centralized scheduling protocol \emph{iOrder} \cite{iOrder} which maximizes channel spatial reuse in interference-oriented scheduling. (Note that iOrder has been shown to outperform scheduling algorithms such as Longest-Queue-First \cite{Ness:DLQF}, GreedyPhysical \cite{interferenceNumber-sch}, and LengthDiversity \cite{geometric-SINR} in maximizing channel spatial reuse.)
    In practice, it is infeasible to apply iOrder to V2V networks due to the highly-varying and large-scale nature of V2V networks. In our study, we take snapshots of network conditions (e.g., topology and channel attenuation) in simulation and then analyze the concurrency in iOrder off line. Therefore, the concurrency in Order represents the limit to which a distributed, online scheduling algorithm may potentially achieve.
}

To understand the effectiveness of the geometric approximation of the PRK model by the gPRK model, we also study a variant of CPS, denoted as OCPS (for Oracle CPS), that is the same as CPS except for its use of the PRK model. In OCPS, we assume that, after a vehicle $R$ has a new estimate for the signal power $P(C,R)$ from another vehicle $C$ to itself, the newly estimated $P(C,R)$ is known to every other potentially interfering vehicle through some oracle without requiring any control signalling packet exchange as we have discussed in Section~\ref{subsec:gPRK}; this way, the costly and sometimes infeasible signal-map-related control signalling overhead is gone, and OCPS can be executed in our simulation environment.
\ifthenelse{\boolean{temp}}
{}
{
???To understand the impact of the design decisions of CPS, we have also studied other variants of CPS (e.g., without agile gPRK model instantiation and vehicle location estimation);
\ifthenelse{\boolean{short}}
{due to the limitation of space, we relegate the detailed discussion to \cite{CPS-TR}. }
{interested readers can find the detailed discussion in Appendix~???.}
}

  \begin{wrapfigure}{r}{.4\linewidth}
    \vspace*{-0.15in}
    \centering
    \includegraphics[width=\linewidth]{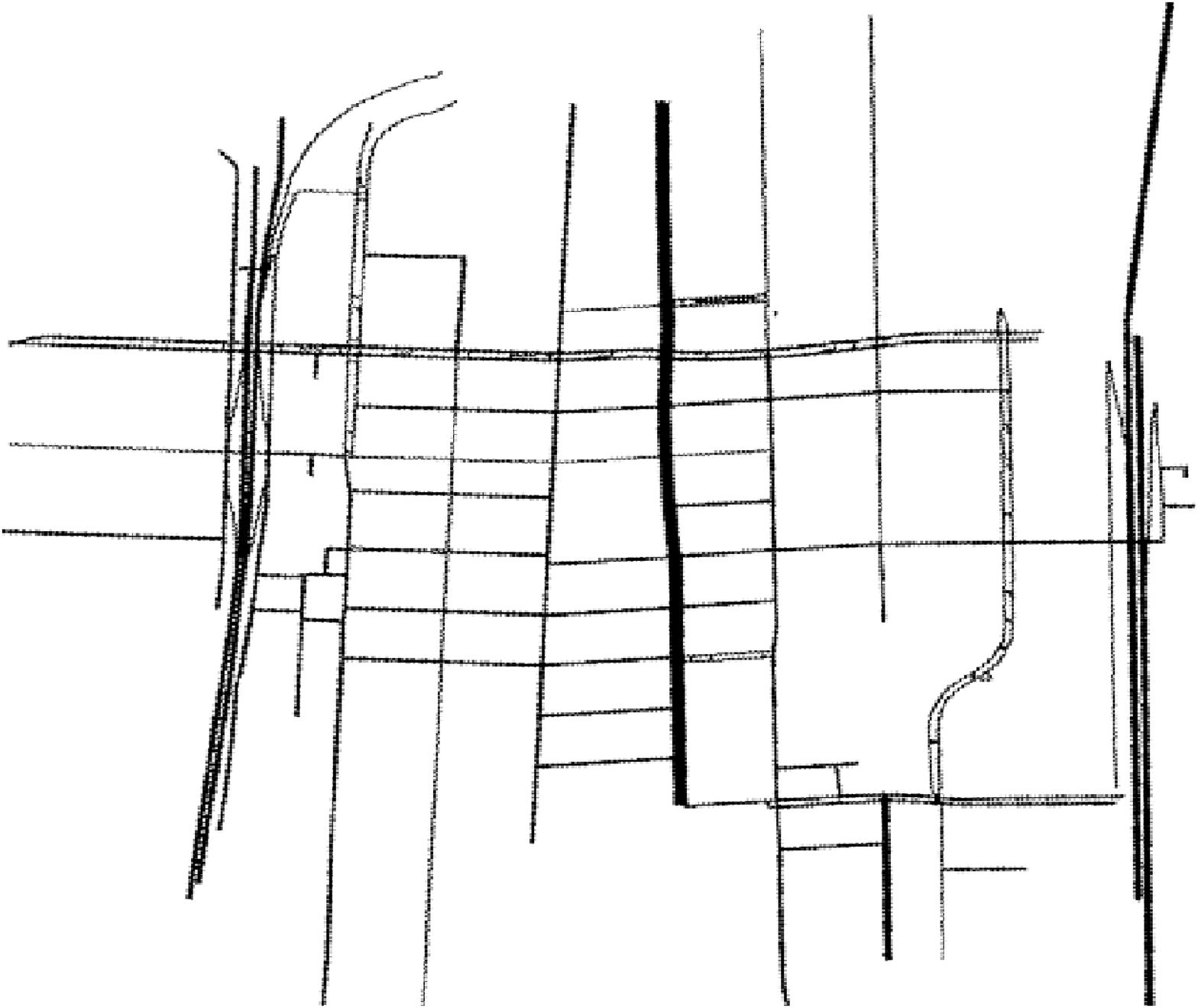}
    \caption{V2V network in Detroit, Michigan, USA} \label{fig:urbanNet}
    \vspace*{-0.15in}
  \end{wrapfigure}
\subHeading{Network settings.}
For understanding protocol behavior in real-world 
settings, we consider an urban network consisting of vehicles in midtown Detroit of Michigan, USA. 
As shown in Figure~\ref{fig:urbanNet}, the urban network consists of freeway I-75 and city roads in midtown Detroit, and it spans an area of $\text{3km} \times \text{3km}$. In the 
network, vehicle speed limits range from 40km/h (i.e., 25mph)
on small city streets to 120km/h (i.e., 75mph) on I-75. 
Our study considers normal vehicle traffic flow conditions, and the average bumper-to-bumper distance 
ranges from one meter to 20 meters. 

We set the desired broadcast communication range as 150 meters and the desired broadcast reliability as 90\%. For protocols that do not use transmission rate and power control (i.e., protocols other than DCC), the transmission rate is set as 6Mbps, and the transmission power is set at a value that ensures that the signal-to-noise ratio (SNR) in the absence of interference is 6dB above the SNR for ensuring 90\% communication reliability for links of length 150 meters.
    Each vehicle transmits a data packet every 100 milliseconds, a frequency needed for many active safety and networked vehicle control applications in V2V networks \cite{Zhang:V2X-survey}.
    The size of each data packet is 1,500 bytes.

We have experimented with other network settings such as on freeways and when the broadcast reliability requirement is 95\%. 
We have observed phenomena similar to what we will present in Section~\ref{subsec:evalResults};
\ifthenelse{\boolean{short}}
{due to the limitation of space, we relegate the detailed discussion to \cite{CPS-TR}. }
{interested readers can find the detailed discussion in the Appendix.}

\subsection{Experimental Results}  \label{subsec:evalResults}

\begin{figure}[!tbp]
\begin{minipage}[t]{0.48\linewidth}
\centering
\includegraphics[width=\figWidthM]{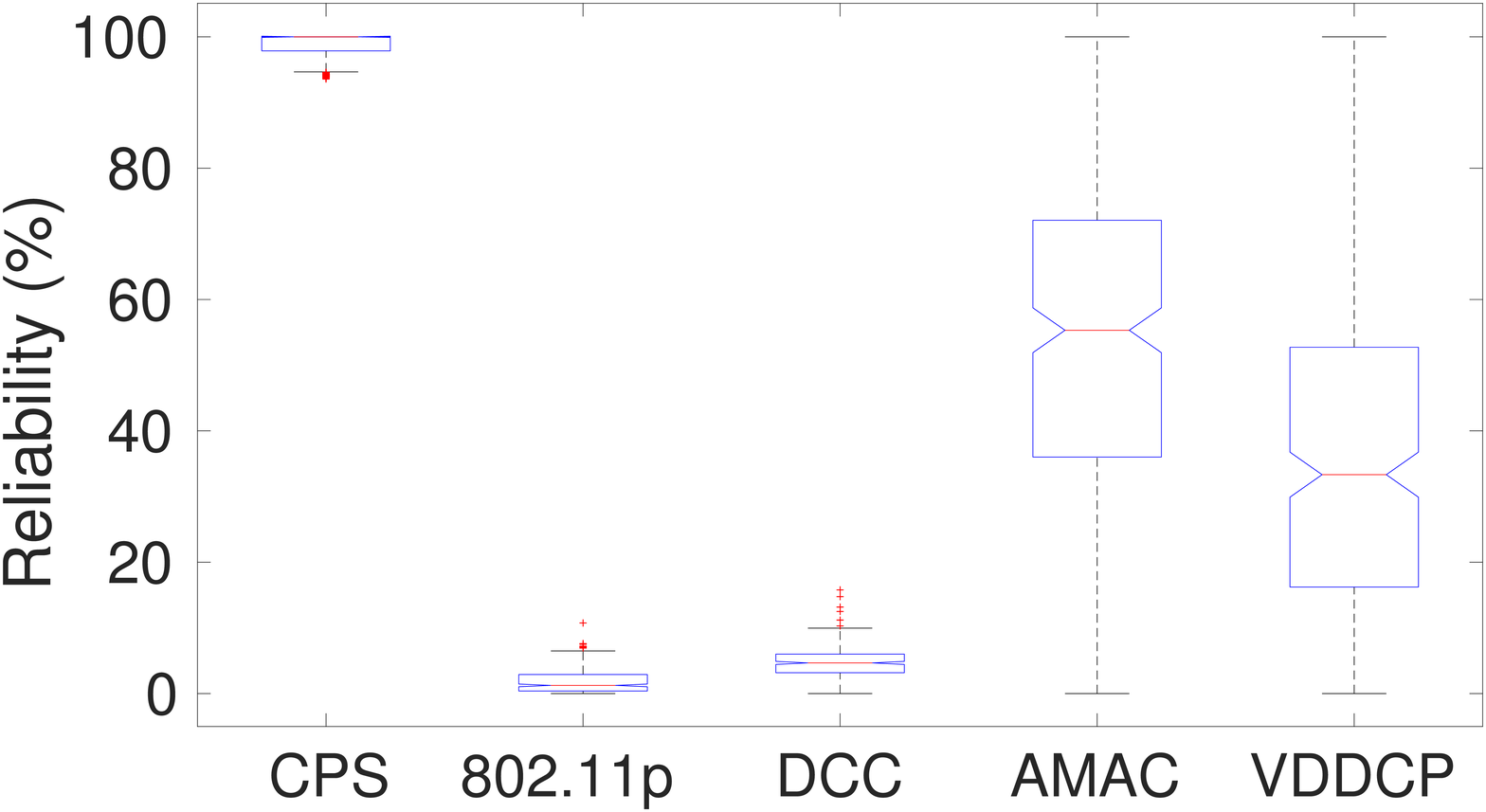}
\caption{Reliability} \label{fig:PDR-urban}
\end{minipage}
\figSpacing
\begin{minipage}[t]{0.48\linewidth}
\centering
\includegraphics[width=\figWidthM]{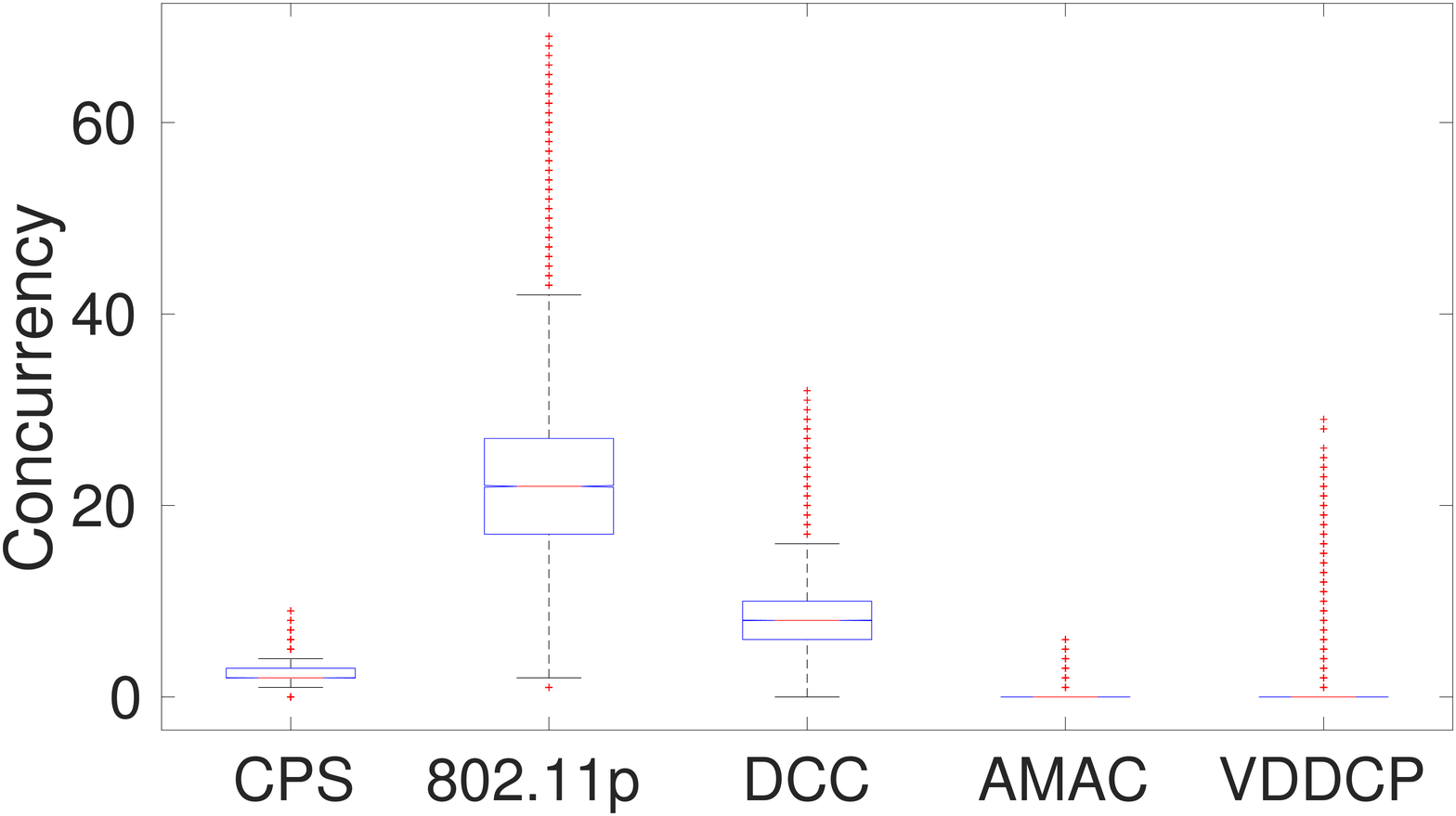}
\caption{Concurrency} \label{fig:concurrency-urban}
\end{minipage}
\\
%
\begin{minipage}[t]{0.48\linewidth}
\centering
\includegraphics[width=\figWidthM]{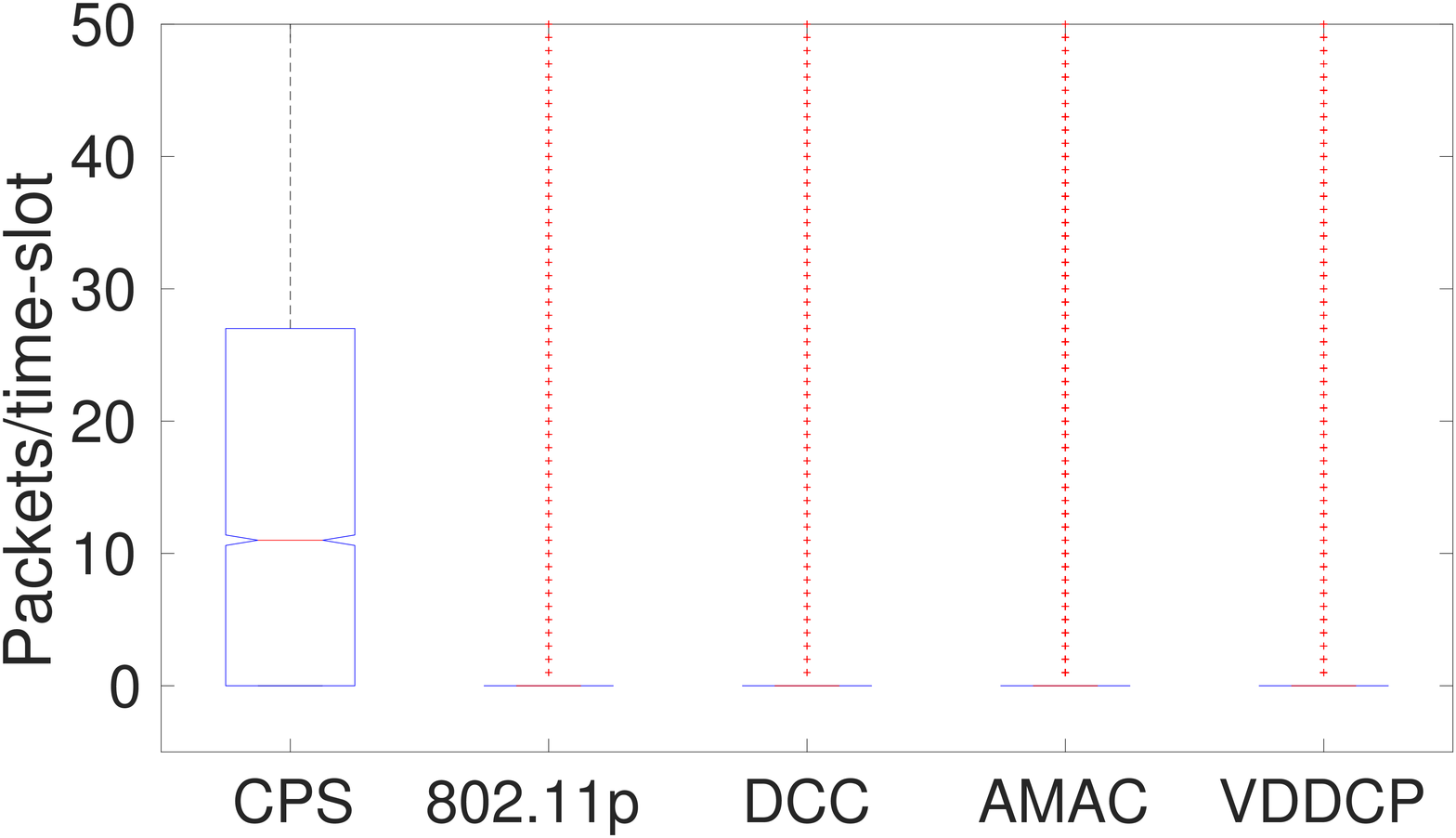}
\caption{Throughput} \label{fig:throughput-urban}
\end{minipage}
\figSpacing
\begin{minipage}[t]{0.48\linewidth}
\centering
\includegraphics[width=\figWidthM]{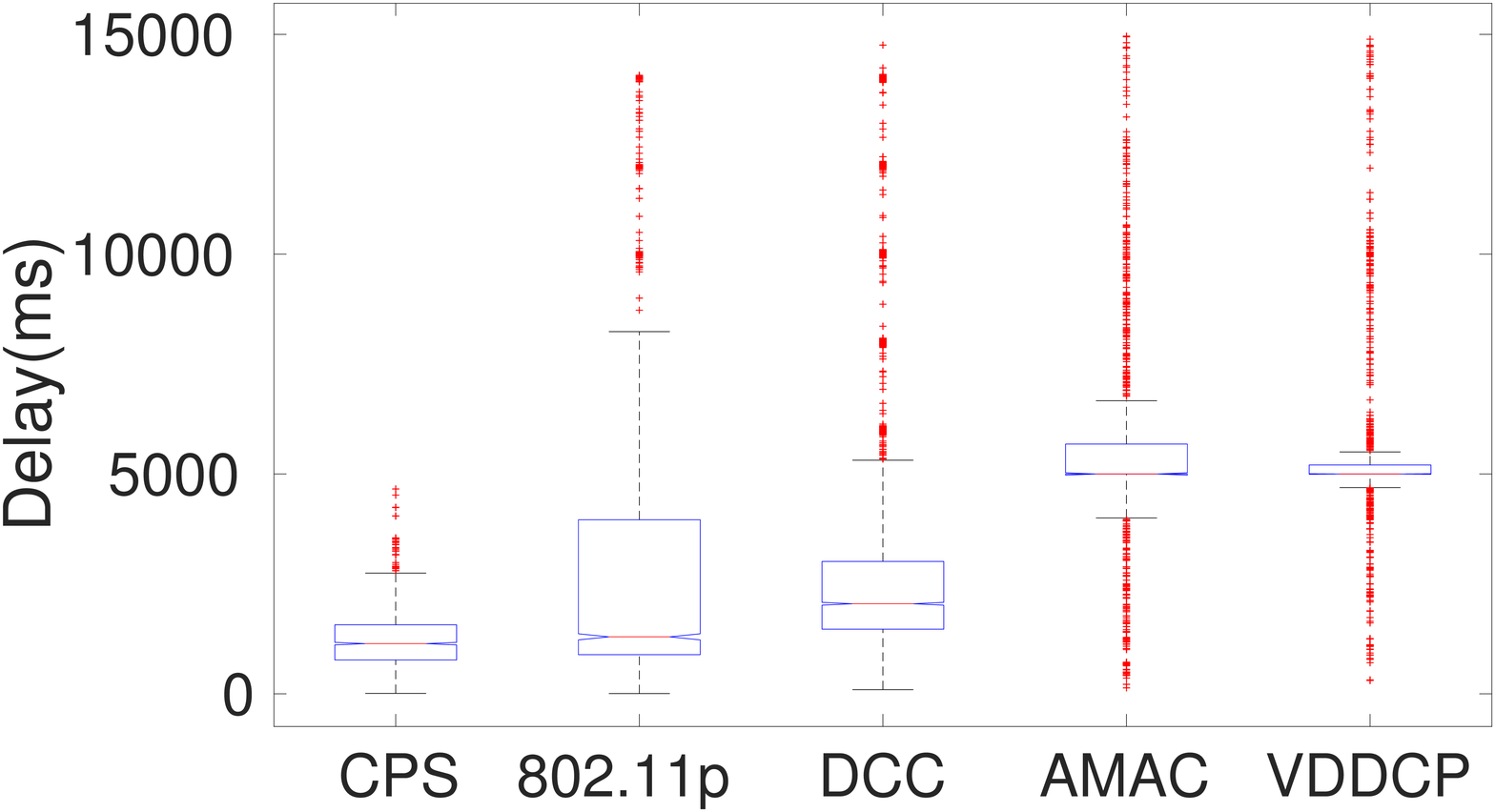}
\caption{Delay} \label{fig:delay-urban}
\end{minipage}
\vspace*{-0.1in}
\end{figure}

\subHeadingS{CPS vs$.$ existing protocols.}
For 
different protocols, Figure~\ref{fig:PDR-urban} shows the boxplot of communication reliability from each vehicle to its receivers, Figure~\ref{fig:concurrency-urban} shows the concurrency (i.e., number of concurrent transmissions) in the network, Figure~\ref{fig:throughput-urban} shows the network throughput that is computed as the number of packets successfully delivered to receivers in every time-slot duration (i.e., 2.5ms), and Figure~\ref{fig:delay-urban} shows the packet delivery delay when packet retransmission is used to ensure the application-required reliability for protocols that would be unable to ensure the application-required reliability otherwise (i.e., protocols other than CPS).

Enabling accurate, agile identification of interference relations among vehicles, our gPRK-based cyber-physical approach to interference modeling and transmission scheduling ensures predictable interference control and application-required broadcast reliability, as shown in Figure~\ref{fig:PDR-urban}.
    Implicitly assuming a protocol interference model and using a contention-based approach to medium access control, 802.11p and DCC do not ensure predictable control of interference and thus do not ensure application-required communication reliability. Through congestion, power, and rate control, DCC improves the reliability of 802.11p, but the broadcast reliability is still quite low in DCC (i.e., being $\sim$6\% in our study).
    Assuming an inaccurate protocol interference model and unable to address the challenge of high vehicle mobility to TDMA scheduling, the TDMA protocols AMAC and VDDCP cannot ensure predictable interference control, and the communication reliability from senders to receivers tend to be quite unpredictable, ranging from very low to very high and varying over time.  In AMAC and VDDCP, the slot reservation tends to be unreliable in the presence of vehicle mobility and inter-vehicle interference, thus the concurrency in AMAC and VDDCP tends to be quite low too, as shown in Figure~\ref{fig:concurrency-urban}. The fact that the reliability is unpredictable while the concurrency is low in AMAC and VDDCP demonstrates the importance of accurately identifying inter-vehicle interference relations in an agile manner in the presence of vehicle mobility, as is accomplished in our CPS framework.

The concurrency in 802.11p and DCC is the highest among all the protocols, but their throughput is quite low due to the low communication reliability in both protocols, as shown in Figures~\ref{fig:throughput-urban} and \ref{fig:PDR-urban}. Due to the low concurrency and the unpredictable, often-low communication reliability in AMAC and VDDCP, the throughput is low in both protocols. Ensuring application-required reliability while maximizing channel spatial reuse, CPS enables significantly higher throughput than other protocols do.

To improve communication reliability, retransmission is needed in other protocols, which significantly increases the communication delay, as shown in Figure~\ref{fig:delay-urban}. The low concurrency and the unpredictable communication reliability in AMAC and VDDCP make their communication delay the largest among all the protocols.

\ifthenelse{\boolean{temp}}
{}
{
  \begin{wrapfigure}{r}{.5\linewidth}
    \vspace*{-0.1in}
    \centering
    \includegraphics[width=\linewidth]{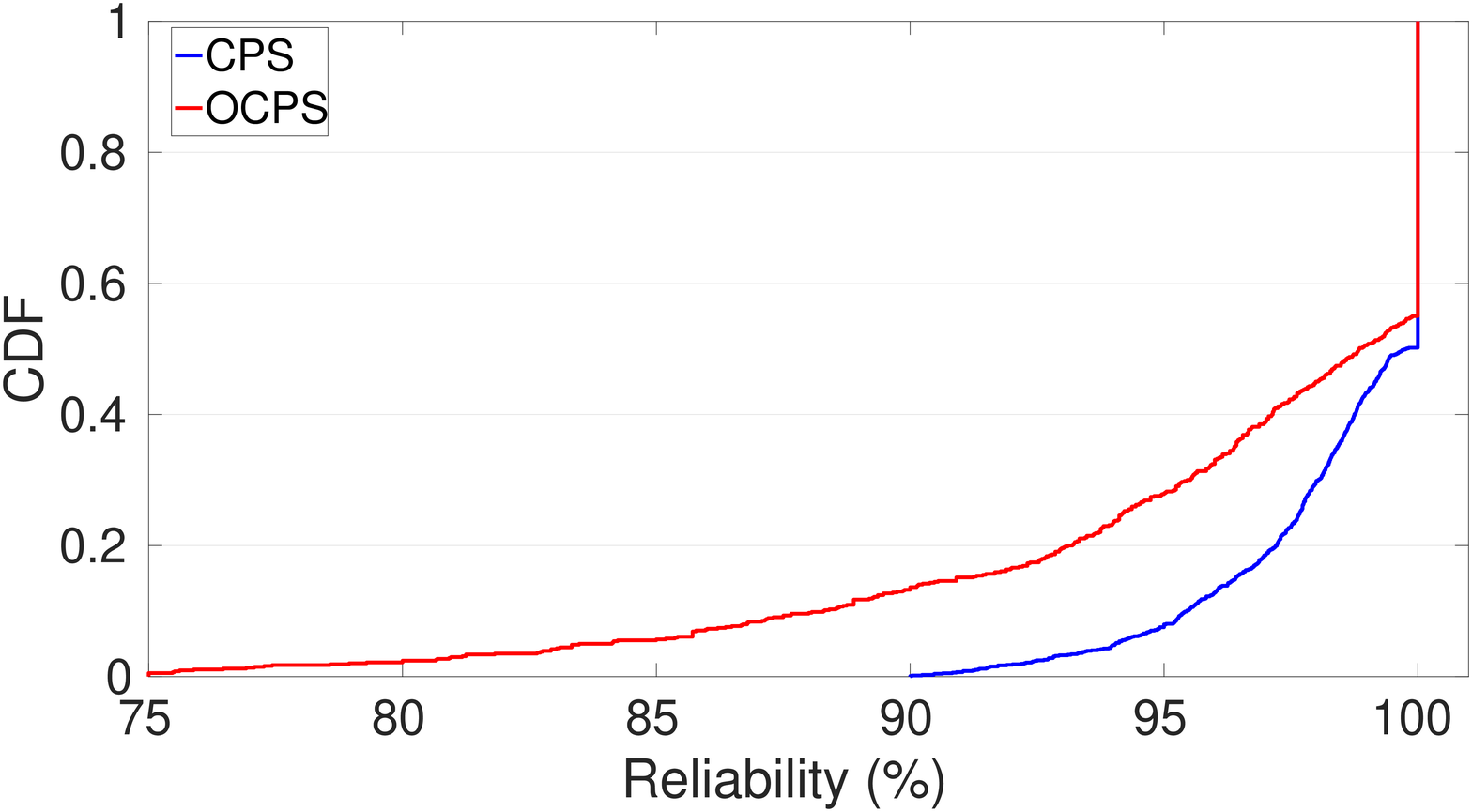}
    \caption{???CPS vs$.$ iOrder} \label{fig:bar-concurrency-PRKS-iOrder}
    \vspace*{-0.1in}
  \end{wrapfigure}
\subHeading{CPS vs$.$ iOrder.}
Figure~\ref{fig:bar-concurrency-PRKS-iOrder} shows the mean concurrency and its 95\% confidence interval in CPS and iOrder for different application requirements on broadcast reliability.
    We see that, despite the highly-varying dynamics in V2V networks and being a distributed protocol, CPS enables a concurrency and spatial reuse statistically equal or close to 
what is enabled by the centralized algorithm iOrder while ensuring the required broadcast reliability. CPS achieves this by ensuring agile, precise identification of mutual interference relations between vehicles and then scheduling maximal, non-interfering concurrent transmissions at each time slot.
    Note that the reason why the concurrency tends to decrease with increasing broadcast reliability requirement in CPS and iOrder is that there exists inherent reliability-throughput tradeoff in channel access scheduling \cite{PRK}.
}

  \begin{wrapfigure}{r}{.5\linewidth}
    \vspace*{-0.1in}
    \centering
    \includegraphics[width=\linewidth]{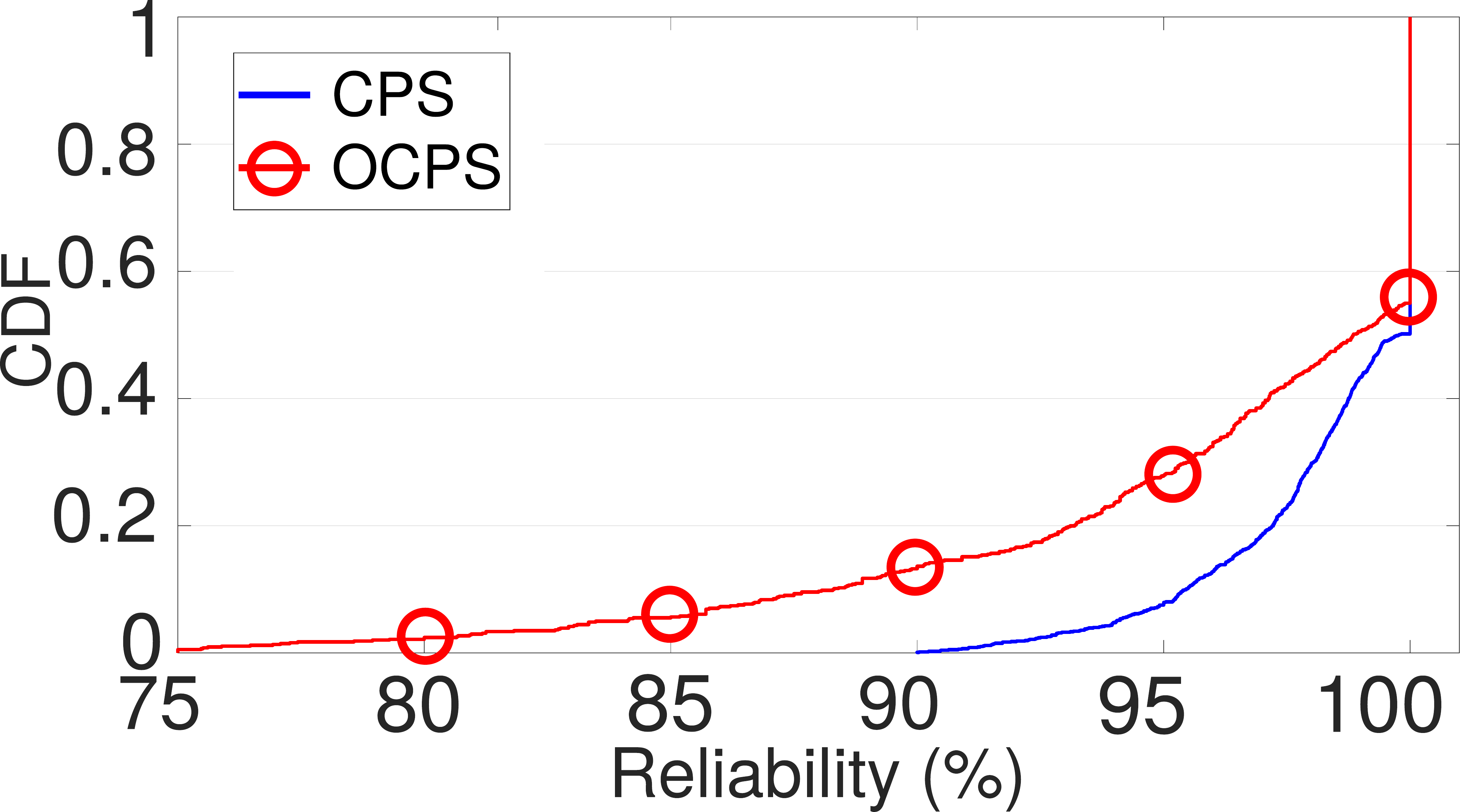}
    \caption{CPS vs$.$ OCPS} \label{fig:CPS-vs-OCPS}
    \vspace*{-0.1in}
  \end{wrapfigure}
\subHeading{CPS vs$.$ OCPS.}
Figure~\ref{fig:CPS-vs-OCPS}
shows the empirical cumulative distribution function (CDF) of the communication reliability from each vehicle to its receivers in CPS and OCPS.
    We see that OCPS achieves a much higher communication reliability than other existing protocols, with the minimum communication reliability being 75\% and the reliability being no less than the required 90\% for about 85\% of the links from a broadcast sender to its receivers.
    Nonetheless, the communication reliability of about 15\% of the links is less than the required 90\% in OCPS, while CPS ensures the required reliability 
    for all the links. The reason for this is because, in OCPS, even though the existence of an oracle addresses the signalling overhead challenge in PRK-based scheduling, it is still difficult to precisely track the highly-dynamic signal power from one vehicle to another in the presence of vehicle mobility, which makes it difficult to precisely track inter-vehicle interference relations and thus difficult to ensure predictable communication reliability.
    In CPS, the gPRK model and the precise tracking of vehicle locations through well-understood vehicle dynamics enable precise tracking of inter-vehicle interference relations and thus enable predictable interference control and predictable communication reliability, showing the benefits of using the geometric approximation of the PRK model in V2V networks.

\section{Related Work} \label{sec:relatedWork}

IEEE 802.11p is a well-studied industry standard specifying the medium access control mechanisms for inter-vehicle communication. Inheriting basic WiFi mechanisms such as CSMA and thus unable to ensure predictable interference control, 802.11p-based solutions do not ensure predictable link reliability 
\cite{Ma:DSRC-reliability-analysis,PRKS}.
	To improve the reliability of inter-vehicle communications, schemes that control information exchange load as well as packet transmission power and rate have been proposed \cite{Xinzhou:VANET-congestion-power-rate-control}. Not addressing the fundamental limitations of CSMA in interference control, these schemes lead to the loss of network throughput and increase in communication delay while still being unable to ensure predictable communication reliability \cite{PRKS}, as we have shown in Section~\ref{sec:eval}.

TDMA schemes \cite{ADHOC-MAC,Valaee:topologyTransparentVANETmac} have also been proposed for inter-vehicle communications. Based on the protocol interference model which is inaccurate and cannot ensure predictable interference control, however, these schemes cannot ensure predictable communication reliability. 
%
Multi-scale schemes have also been proposed to first allocate non-overlapping sets of time slots to different roads and then let vehicles on each road compete for channel access in a TDMA manner \cite{Kumar:VANET-structure-MAC,Zhuang:VeMAC,Lai:regionTDMA}. Assuming a protocol interference model in both road-level scheduling and vehicle-level scheduling, however, these schemes do not ensure predictable communication reliability. 
Schemes have also been proposed to first partition space into geographic regions such as rectangles or hexagons and then schedule transmissions based on geographic regions \cite{Rahul:LDMA,Welch:vanet-MAC}. Assuming a protocol interference model, however, these schemes do not ensure predictable communication reliability either. 
Resource allocation mechanisms have also been proposed to improve communication throughput between vehicles as well as between vehicles and transportation infrastructures \cite{Yao:v2v:v2I}. Focusing on network throughput, these work do not consider ensuring predictable, controllable reliability in vehicular communication, and, due to throughput-reliability tradeoff \cite{PRK}, the high throughput usually comes at the cost of low communication reliability.

\section{Concluding Remarks} \label{sec:concludingRemarks}

For predictable reliability of inter-vehicle communications, we formulate and apply the gPRK interference model to predictable interference control in V2V networks. Our approach to gPRK-based interference modeling effectively leverages cyber-physical structures of V2V 
\ifthenelse{\boolean{short}}
{networks.}
{networks, particularly, spatiotemporal interference correlation, correlated receiver ER adaptation, and set-cover-based control signaling in the cyber-domain, and vehicle locations as well as macro- and micro-scopic vehicle dynamics in the physical domain.}
Based on the cyber-physical, gPRK-based approach to interference modeling, our Cyber-Physical Scheduling (CPS) framework ensures predictable reliability of inter-vehicle communications.
    Ensuring predictable interference control and communication reliability in the presence of vehicle mobility, our cyber-physical approach to interference modeling and data transmission  scheduling is expected to enable the development of mechanisms for predictable timeliness, throughput, and their tradeoff with reliability in inter-vehicle communications. While our focus in this study is on inter-vehicle communications, the basic methodologies can be extended to enable predictable communication reliability between vehicles and transportation infrastructures such as traffic lights. These are future directions worth pursuing.

\ifthenelse{\boolean{anony} \OR \NOT \boolean{piwi} }
{}
{\input{ack} }

{\small
\bibliographystyle{abbrv}
\bibliography{references}
}

\ifthenelse{\boolean{short}}
{
\newpage

}
{
\input{appendix-eval}
}


\end{document}